\documentclass[]{spie}  

\usepackage{amsmath,amsfonts,amssymb}
\usepackage{graphicx}
\usepackage[colorlinks=true, allcolors=blue]{hyperref}

\usepackage{float}
\usepackage{caption}
\usepackage{subcaption}
\usepackage{array}

\title{Crank-rocker optical fiber mode scrambler prototype for the GMT-Consortium Large Earth Finder (G-CLEF)}

\author[a,b,c]{Matthew C. H. Leung}
\author[a,c]{Colby Jurgenson}
\author[a,b,c]{Andrew Szentgyorgyi}
\author[a,c]{William Podgorski}
\author[a,c]{Mark Mueller}
\author[d]{Yahel Sofer Rimalt}
\author[a,c]{Joseph Zajac}
\author[a,c]{Cem Onyuksel}
\author[a,c]{Daniel Durusky}
\author[a,c]{Peter Doherty}

\affil[a]{Center for Astrophysics \textbar{} Harvard \& Smithsonian, 60 Garden St, Cambridge, MA 02138, USA}
\affil[b]{Department of Astronomy, Harvard University, 60 Garden St, Cambridge, MA 02138, USA}
\affil[c]{Smithsonian Astrophysical Observatory, 100 Acorn Park Dr, Cambridge, MA 02140, USA}
\affil[d]{Weizmann Institute of Science, 234 Herzl St, Rehovot, 7610001, Israel}

\authorinfo{Further author information: (Send correspondence to M.C.H. Leung)\\M.C.H. Leung: E-mail: matthew.leung@cfa.harvard.edu \\  C. Jurgenson: E-mail: colby.jurgenson@cfa.harvard.edu}

\begin{document} 
\maketitle

\begin{abstract}
When coherent light propagates through a multimode optical fiber, the modes interfere at the fiber exit boundary, producing a high-contrast speckle interference pattern called modal noise. This non-uniform interference pattern introduces systematic errors in fiber-fed precision radial velocity (RV) spectrographs which are detrimental to exoplanet mass measurement. Modal noise can be mitigated by a device called a fiber mode scrambler or fiber agitator, which dynamically perturbs the fiber to change the interference pattern over time, smoothing it over long exposures. In this paper, we present a prototype optical fiber mode scrambler based on a four-bar linkage crank-rocker mechanism, developed for the GMT-Consortium Large Earth Finder (G-CLEF). G-CLEF is a fiber-fed, high-resolution, precision RV spectrograph for the Magellan Clay Telescope and Giant Magellan Telescope (GMT). To support this effort, we developed a fiber testing setup capable of imaging the near-field and far-field output of fibers and measuring focal ratio degradation. We designed, built, and tested the mode scrambler, using our setup, on step-index multimode optical fibers with various shapes, including octagonal, square, and rectangular core cross-sections. We developed custom software utilizing alpha shapes to identify the boundary of an arbitrarily shaped fiber and to compute a signal-to-noise ratio metric for quantifying modal noise. We investigated the effects of different mode scrambler parameters, such as agitation frequency, on mitigating modal noise. Our results offer valuable insights into optimizing fiber mode scrambling for precision RV spectrographs.
\end{abstract}

\keywords{optical fiber mode scrambler, optical fiber agitator, modal noise, precision radial velocity, alpha shape, G-CLEF, GMT-Consortium Large Earth Finder}


\section{Introduction}\label{sec:intro}

The radial velocity (RV) method\cite{Mayor1995} has been used to discover more than a thousand exoplanets to date. This method works by measuring the Doppler shift of the spectral lines of a star, due to the star's line-of-sight motion caused by the gravitational influence of orbiting planets. One of the highest-priority programs in contemporary astrophysics\cite{NAdecadal} is to discover rocky Earth-mass exoplanets orbiting in the habitable zone of Sun-like stars; or in other words, to find ``Earth 2.0''. This would require a RV precision of $\sim$10 cm/s, which is the line-of-sight reflex motion of a solar-mass star induced by an Earth-mass planet orbiting at 1 AU away, an order of magnitude improvement in measurement precision over most conventional high-resolution astronomical spectrographs. This goal imposes challenging requirements on spectrograph stability and precision.

To tackle this challenge, we are developing the GMT-Consortium Large Earth Finder (G-CLEF\cite{Szentgyorgyi2024,Mueller2022,Szentgyorgyi2018,Szentgyorgyi2016}), which is a fiber-fed, high-resolution, precision RV, optical echelle spectrograph specifically designed to search for an ``Earth 2.0''. G-CLEF will operate on the 6.5 m Magellan Clay Telescope (G@M\cite{Jurgenson2023,Sofer-Rimalt2024}), before being moved to the future 25.4 m Giant Magellan Telescope (GMT). At the time of writing, G-CLEF is currently under assembly\cite{Jurgenson2024, Mueller2024, Leung2024, Sofer-Rimalt2024}, with first light at Magellan anticipated in 2027 and commissioning at the GMT anticipated in the 2030s.

Most astronomical spectrographs fed by multimode optical fiber, like G-CLEF, suffer from a phenomenon called modal noise\cite{Epworth1979}. An optical fiber is a waveguide, and light propagates through a multimode optical fiber in a finite integer number of transverse electromagnetic modes. These modes are eigensolutions of the Helmholtz equation on the fiber cross section \cite{SalehTeich}. When coherent light propagates through a multimode optical fiber, the modes can interfere to produce a high-contrast speckle interference pattern at the fiber exit boundary; this is modal noise. Modal noise severely decreases the signal-to-noise ratio of the extracted velocity from spectra taken by high-resolution spectrographs \cite{Petersburg2018,Sablowski2016,Baudrand2001}. The non-uniformity of the interference pattern affects the illumination of the spectrograph, leading to systematic errors which can compromise wavelength calibration \cite{Sirk2018,Frank2018}. This introduces drift\cite{Blackman2020} in the extracted velocity, ultimately resulting in lower precision of exoplanet mass measurements. However, if the fiber is moved, then the interference pattern will change because the shape of the waveguide will change. Hence, by mechanically agitating\cite{Baudrand2001} the fiber, the interference pattern will change over time and be smoothed out over long exposures. A device used to mitigate modal noise is called a ``fiber mode scrambler'' or a ``fiber agitator''.

While most optical fiber mode scramblers mechanically agitate the fiber to mitigate modal noise, the generality of mode scramblers for astronomical spectrographs in the literature are limited, because they are often specific to a given instrument\cite{Sablowski2016}. Different mode scramblers in the literature mechanically agitate an optical fiber in different ways, for example by shaking\cite{Frank2018,Sablowski2016}, bending\cite{Ishizuka2018}, rotating\cite{Petersburg2018}, or twisting\cite{Ishizuka2018}. In this work, we investigate a mode scrambler based on a four-bar linkage crank-rocker mechanism, which aims to unify these different methods of mechanical agitation. Our crank-rocker mechanism induces compound motion that simultaneously bends, rotates, and oscillates an optical fiber, combining the strengths of different agitation methods into a single device to achieve more effective mode scrambling. Our crank-rocker mode scrambler was a prototype developed for G-CLEF.

An outline of this paper is as follows. In Section \ref{sec:FCS}, we discuss a multi-purpose fiber characterization station we built to image the near-field and far-field output of optical fibers and to measure optical fiber focal ratio degradation (FRD). This setup was used to measure mode scrambling. In Section \ref{sec:MS_design}, we discuss the design of our four-bar linkage crank-rocker mode scrambler. In Section \ref{sec:analysis}, we discuss our image analysis procedure and how we quantified mode scrambling. In Section \ref{sec:results}, we present the performance of our mode scrambler, tested on several step-index multimode optical fibers with different cross-sectional shapes.

\section{Fiber Characterization Station}\label{sec:FCS}

\subsection{Overview}\label{sec:FCS_overview}

In this section, we discuss the fiber characterization station (FCS) we built at the Smithsonian Astrophysical Observatory (SAO) Cambridge Discovery Park (CDP) laboratories to test optical fibers agitated by our mode scrambler. Enough detail is provided so that a reader can reproduce our setup. The FCS is a multi-purpose platform that is used for characterizing several properties of optical fibers. While the primary focus of this paper is on mode scrambling, our FCS can also support measurements of scrambling gain\cite{Avila2006,Halverson2015} (also known as optical scrambling\cite{Hunter1992}), a distinct but related concept that describes the fiber's ability to decouple the output illumination distribution from the input illumination distribution. Our FCS has four goals, which are to:
\begin{enumerate}
    \itemsep0em 
    \item Image fiber input face (to ensure proper light injection location and alignment)
    \item Image fiber near-field (fiber output face, for mode scrambling measurements)
    \item Image fiber far-field (collimated output of fiber)
    \item Measure fiber output power, throughput, and focal ratio degradation (FRD)
\end{enumerate}
The first two goals are directly related to mode scrambling. The latter two goals allow for measurements of scrambling gain. The design of our FCS setup at SAO CDP was inspired by a FCS setup at Yale University \cite{Petersburg2018,Tuttle2020}. Our FCS design consists of four arms:
\begin{enumerate}
    \itemsep0em 
    \item Pre-injection arm, to inject light into the fiber (Section \ref{sec:pre-injection_arm})
    \item Injection imaging arm, to image the fiber input face (Section \ref{sec:injection_imaging_arm})
    \item Near-field arm, to image the fiber near-field (Section \ref{sec:near_field_arm})
    \item Far-field arm, to image the fiber far-field (Section \ref{sec:far_field_arm})
\end{enumerate}

\begin{figure}[h]
    \centering
    \includegraphics[width=0.8\textwidth]{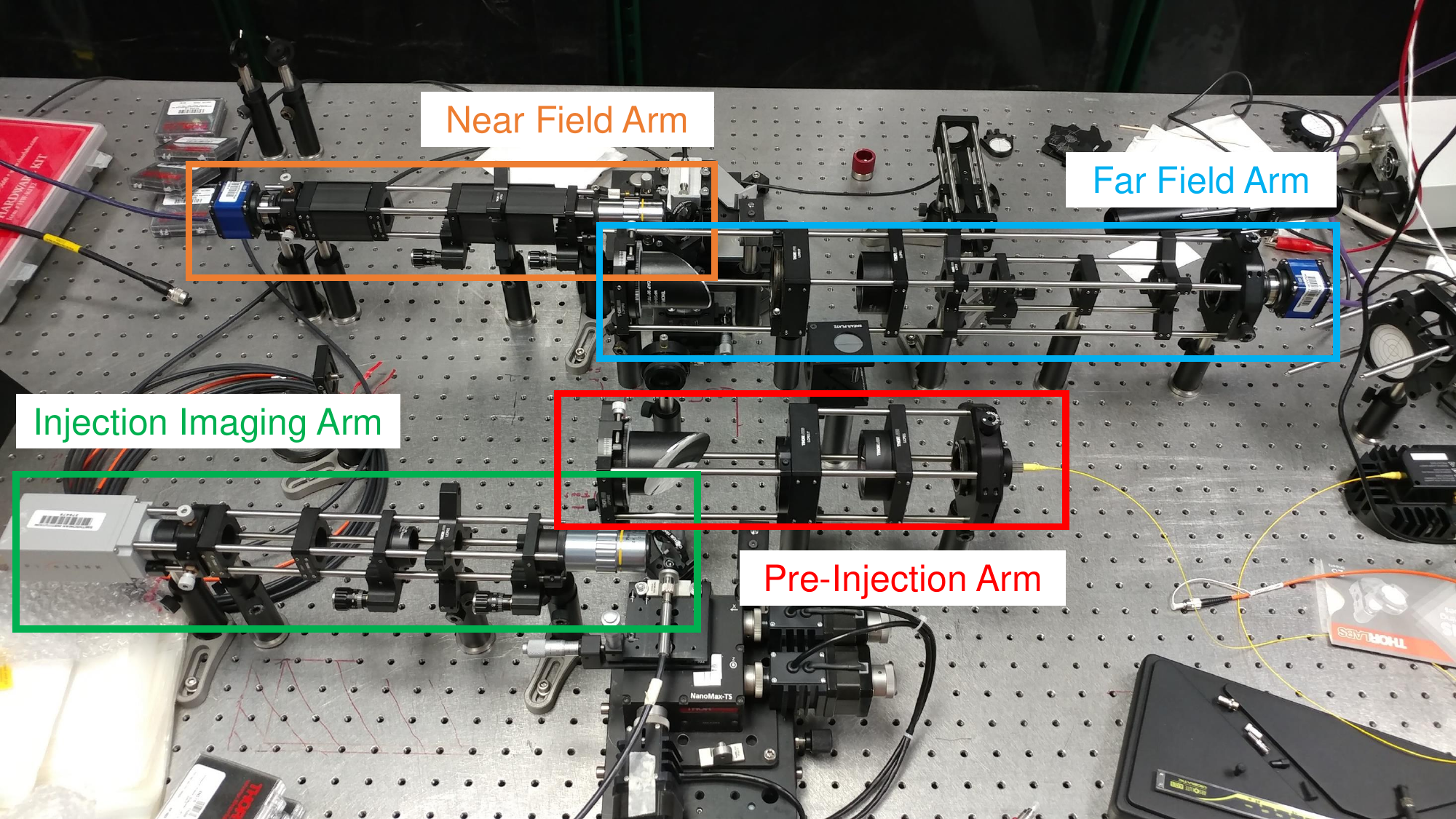}
    \caption{Fiber characterization station, with the four arms labeled}
    \label{fig:FCS_overview}
\end{figure}

Figure \ref{fig:FCS_overview} shows the FCS and its four arms. The FCS has two configurations: one in which light is injected through the fiber, and one in which the fiber is bypassed for throughput measurement. 

\subsubsection{Configuration 1: Through the Fiber}\label{sec:FCS_config1}

\begin{figure}[bh]
    \centering
    \includegraphics[width=0.8\textwidth]{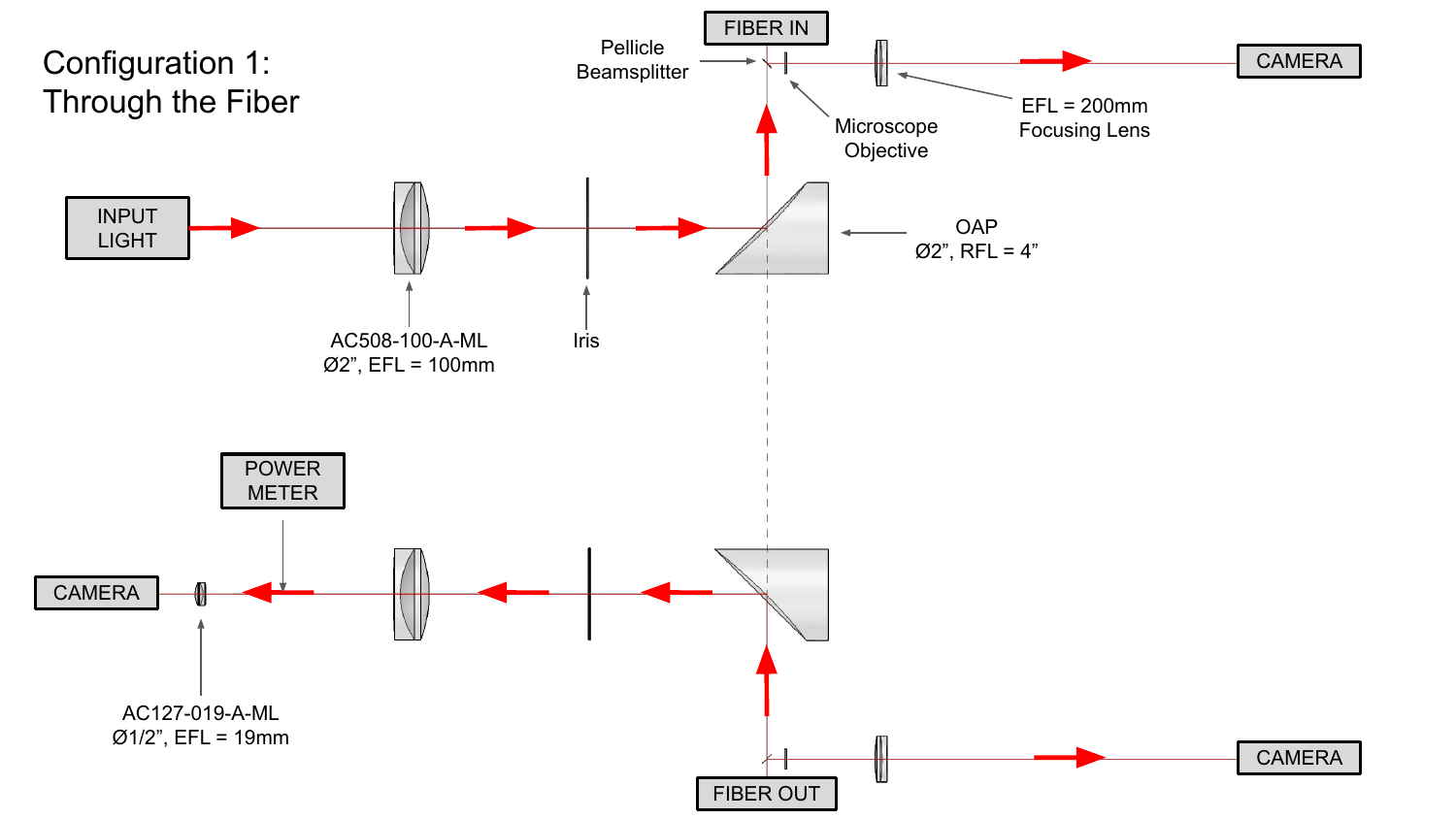}
    \caption{Configuration 1 of the FCS. Note that the components are presented here such that this figure is a $180^\circ$ rotated version of Figure \ref{fig:FCS_overview}.}
    \label{fig:FCS_config1}
\end{figure}

Figure \ref{fig:FCS_config1} shows the first configuration of the FCS. In the first configuration, light is injected into the fiber. Firstly, input light from a certain light source is collimated by an achromatic doublet lens and is then focused by a $90^\circ$ off-axis parabolic mirror (OAP) onto the input face of the test fiber. We chose to use OAPs in our FCS because of their achromatic response and their ability to redirect light. Between the lens and the OAP is an iris diaphragm. This is the pre-injection arm. The pre-injection arm basically reimages the input light onto the fiber input face. This reimaging is nearly one-to-one.

Light is also reflected from the fiber input face. This reflected light is then reflected by a pellicle beamsplitter into the injection imaging arm, which is a microscope. After reflection from the pellicle beamsplitter, the light is collimated by an infinite conjugate microscope objective and is then focused by an achromatic doublet lens onto a detector. This arm of the FCS images the fiber input face, to ensure that the light source is correctly injected into the fiber.

The light which is injected into the fiber exits the fiber on the other side of the FCS, which is nearly a mirror copy of the injection side. The output light from the fiber is then split into two directions by a pellicle beamsplitter. Some of the light is reflected by the pellicle beamsplitter into the near-field arm, which is nearly identical to the injection imaging arm. This reflected light is then collimated by an infinite conjugate microscope objective and is then focused by an achromatic doublet lens onto a detector. The near-field arm images the output face of the fiber, hence imaging the near-field of the fiber. The only difference between the near-field arm and the injection imaging arm is the detector used.

The rest of the light which is not reflected by the pellicle beamsplitter instead passes through the beamsplitter and is collimated by a $90^\circ$ OAP. The collimated beam is refocused by an achromatic doublet lens. Between the OAP and the lens is an iris diaphragm. At the focus of the lens is a power meter, which is removable. The power meter is used to measure the output power of the fiber, for FRD measurements. If the power meter is not in place, then the focused beam expands and is recollimated by another achromatic doublet lens. The collimated beam then hits a detector, which images the far-field of the fiber. This is the far-field arm.


\subsubsection{Configuration 2: Throughput Measurement}\label{sec:FCS_config2}
\begin{figure}[h]
    \centering
    \includegraphics[width=0.8\textwidth]{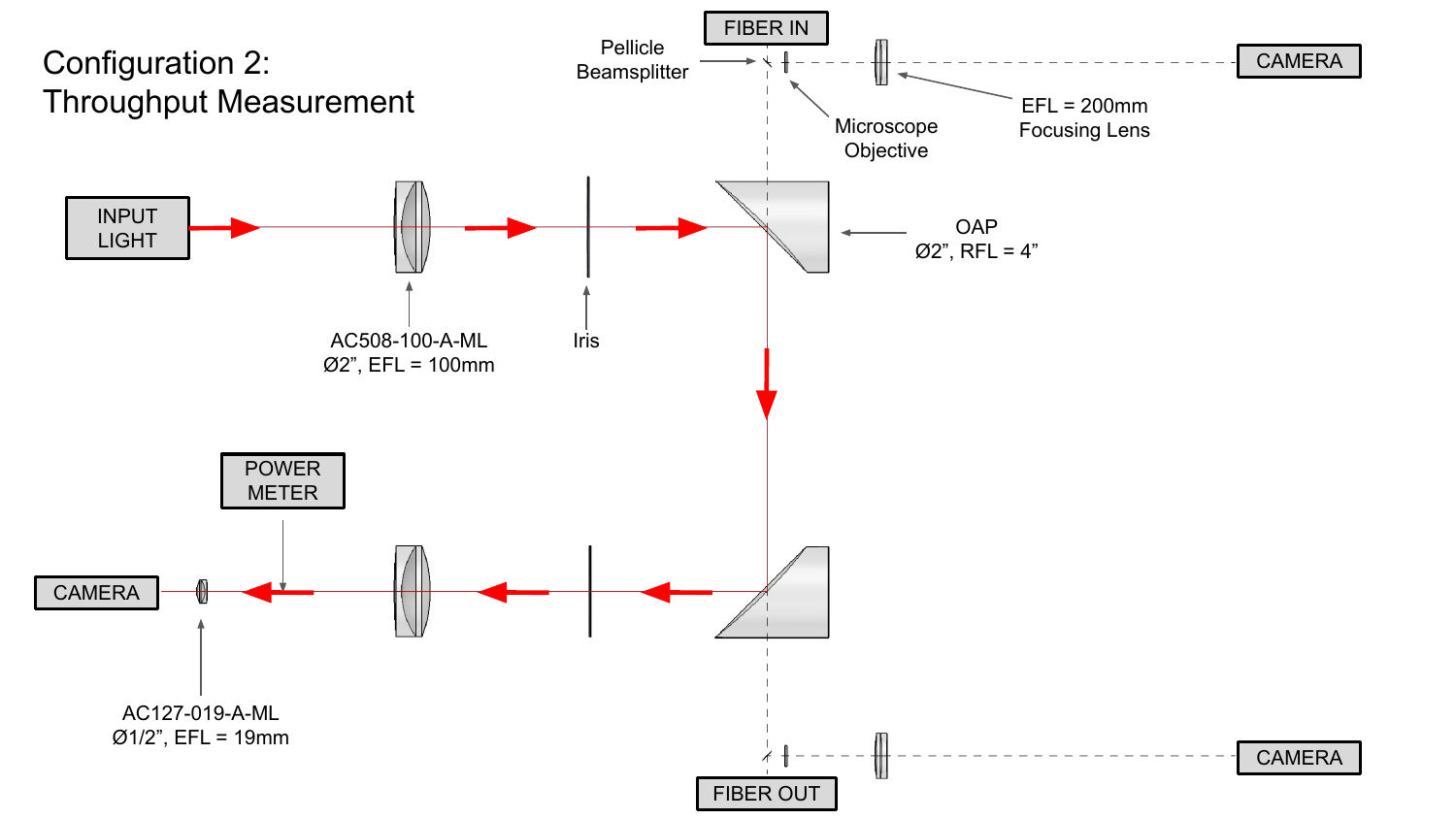}
    \caption{Configuration 2 of the FCS. Notice that compared to Figure \ref{fig:FCS_config1}, both OAPs are rotated by $180^\circ$.}
    \label{fig:FCS_config2}
\end{figure}

Figure \ref{fig:FCS_config2} shows the second configuration of the FCS. In the second configuration, the OAPs are rotated $180^\circ$ to bypass the fiber. Light from the injection arm instead directly goes to the far-field arm. The purpose of this configuration is to allow for the throughput of the FCS system to be accounted for when taking measurements of the output power from the fiber. The lenses and mirrors in the FCS will not have perfect transmission and reflection respectively, and so there will be some loss due to the components in the FCS. The loss due to only the fiber is the difference between the power measured in this configuration and the power measured in  \mbox{configuration 1}.

In addition to measuring loss in the fiber, this configuration is necessary for FRD measurements. This is done by adjusting the iris diaphragms. Let the irises in the injection arm and far-field arm be called iris 1 and iris 2 respectively. Suppose that we want to measure the FRD for a particular starting focal ratio $N$. To do this, we first adjust iris 1 so that the focal ratio of the input beam being injected into the fiber is $N$. We then measure the power in configuration 2, in which the fiber is bypassed. After this value is recorded, we switch to configuration 1, in which light is injected into the fiber, and measure the power. Due to FRD, the power of the output beam of the fiber will be less than the power measured in configuration 2. We then close iris 2 until the measured power is close (e.g., $90\%$) to the measured power in configuration 2. Suppose to do this, we need a focal ratio of $N'$ for iris 2. Hence, the FRD is $N'-N$.


\subsection{Pre-Injection Arm}\label{sec:pre-injection_arm}

\begin{figure}[h]
    \centering
    \includegraphics[width=0.8\textwidth]{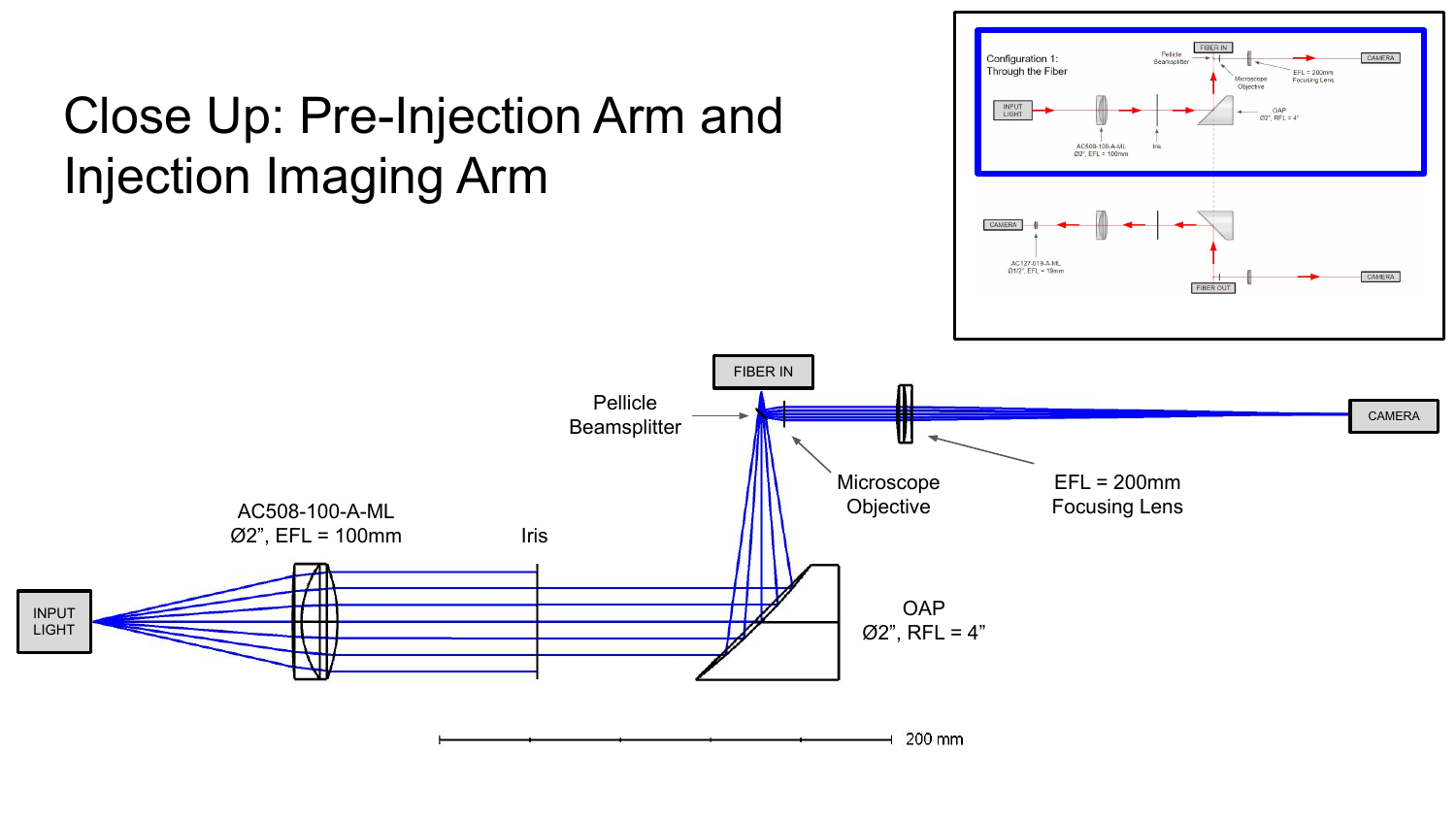}
    \caption{Zemax OpticStudio model of the pre-injection arm and injection imaging arm. The inset at the top right shows the location of these components in Figure \ref{fig:FCS_config1}.}
    \label{fig:FCS_zmx_preinject_injectimg}
\end{figure}

\begin{figure}[h]
    \centering
    \includegraphics[width=0.75\textwidth]{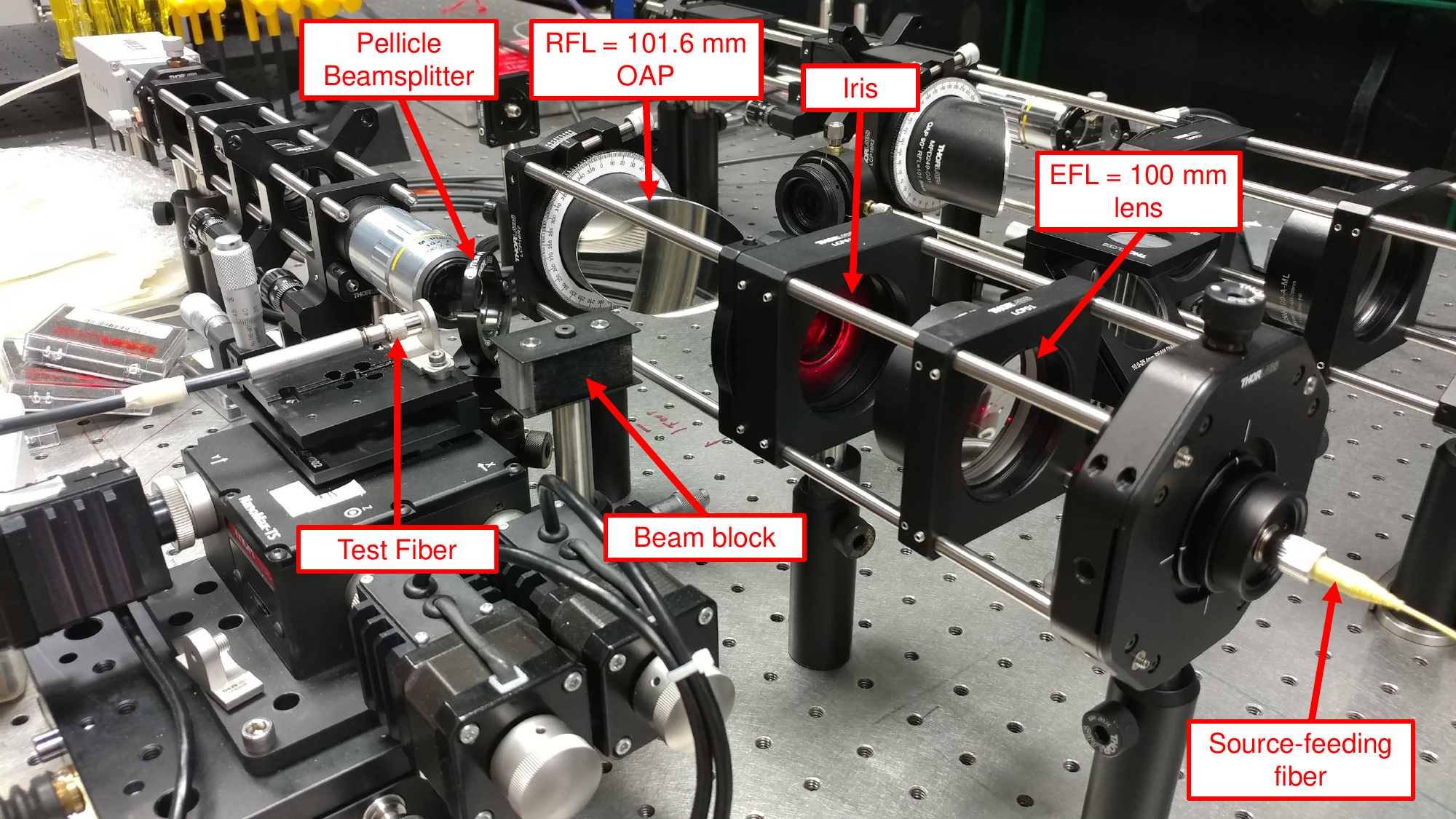}
    \caption{Pre-injection arm, with components labeled}
    \label{fig:FCS_preinject}
\end{figure}

The pre-injection arm, shown in Figures \ref{fig:FCS_zmx_preinject_injectimg} and \ref{fig:FCS_preinject}, consists of all the optical components before light is injected into the test fiber. Its purpose is to inject light into the fiber from a light source. Several light sources were used with the FCS: a Thorlabs LPS-635-FC 635 nm laser diode, a Hamamatsu Photonics Energetiq Laser Driven Light Source (LDLS), and a Thorlabs white light mounted LED. All of these sources were fed to the FCS via single mode optical fiber.

The pre-injection arm uses a 60 mm cage system from Thorlabs. The output end of the source-feeding fiber is connected to the pre-injection arm by a Thorlabs SM1-threaded fiber adapter, which is held in place by an XYZ translation mount (Thorlabs model CXYZ1). The degrees of freedom offered by the XYZ translation mount allow the input light source position to be adjusted for alignment purposes. In particular, XY translation capability is needed for astigmatism correction during alignment. The cone of light outputted from the source-feeding fiber is collimated by a Ø$2''$ 100 mm effective focal length (EFL) achromatic doublet lens (Thorlabs model AC508-100-A-ML).

The collimated beam of light then passes through an iris diaphragm, which is mounted on the cage system. Afterwards, the collimated beam is focused by a Ø$2''$ $90^\circ$ OAP with a reflected focal length (RFL) of 101.6 mm (Thorlabs model MPD249-G01) onto the input face of the test fiber. The achromatic doublet lens and OAP form a nearly one-to-one relay system due to their focal lengths; output light from the source-feeding fiber is reimaged nearly one-to-one onto the input face of the test fiber. The OAP is held by a rotation mount (Thorlabs model LCP16R2), to allow for the OAP to be rotated between the two configurations. The test fiber is held in place by a tip-tilt stage mounted on a Thorlabs NanoMax multi-axis flexure stage, which allows the test fiber input position to be adjusted with precision. We ensure that the image injected into the test fiber is smaller in size compared to the test fiber, by using a single mode fiber to feed the light sources.

Between the OAP and the test fiber is a Ø$1''$ pellicle beamsplitter which directs reflected light from the fiber input face towards the injection imaging arm. Instead of other types of beamsplitters, a pellicle beamsplitter was selected because it does not cause ghosting effects and beam deviation. The pellicle beamsplitter has a reflection-transmission ratio of 45:55 (Thorlabs model BP145B1), and is oriented at an angle of $45^\circ$ with respect to the path in which the focusing beam (from the OAP) is traveling. As a result of this pellicle beamsplitter, some of the incoming light which is being focused by the OAP will be reflected in the direction directly opposite of the injection imaging arm, instead of being injected into the test fiber. A beam block was added in order to absorb the reflected incoming light.

\subsection{Injection Imaging Arm}\label{sec:injection_imaging_arm}

\begin{figure}[h]
    \centering
    \includegraphics[width=0.8\textwidth]{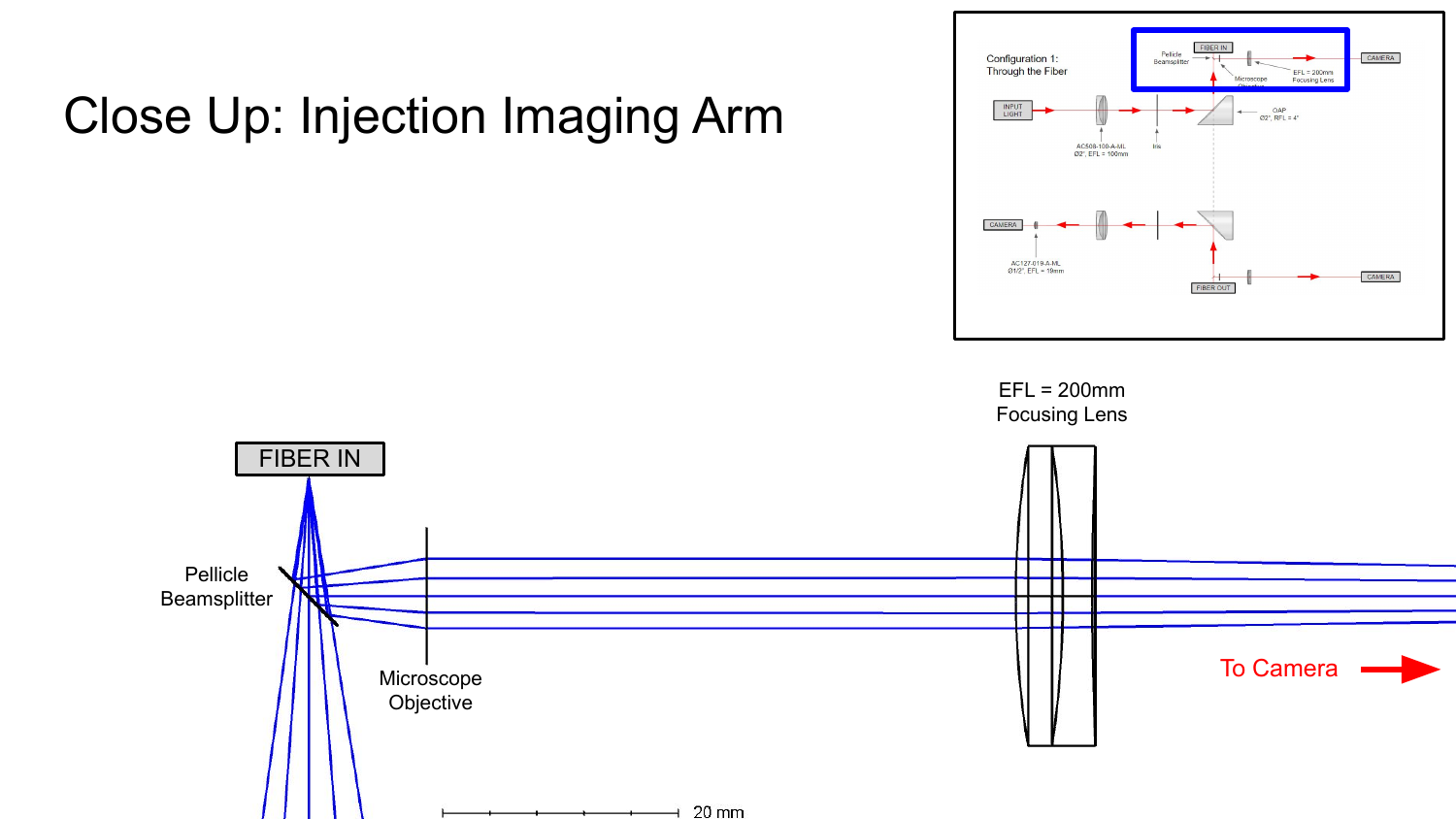}
    \caption{Zemax OpticStudio model of part of the injection imaging arm, from the pellicle beamsplitter to the 200 mm EFL lens. Note that the microscope objective is modeled by a paraxial lens because an actual model could not be found. The inset at the top right shows the location of these components in Figure \ref{fig:FCS_config1}.}
    \label{fig:FCS_zmx_injectimg}
\end{figure}

\begin{figure}[h]
    \centering
    \includegraphics[width=0.8\textwidth]{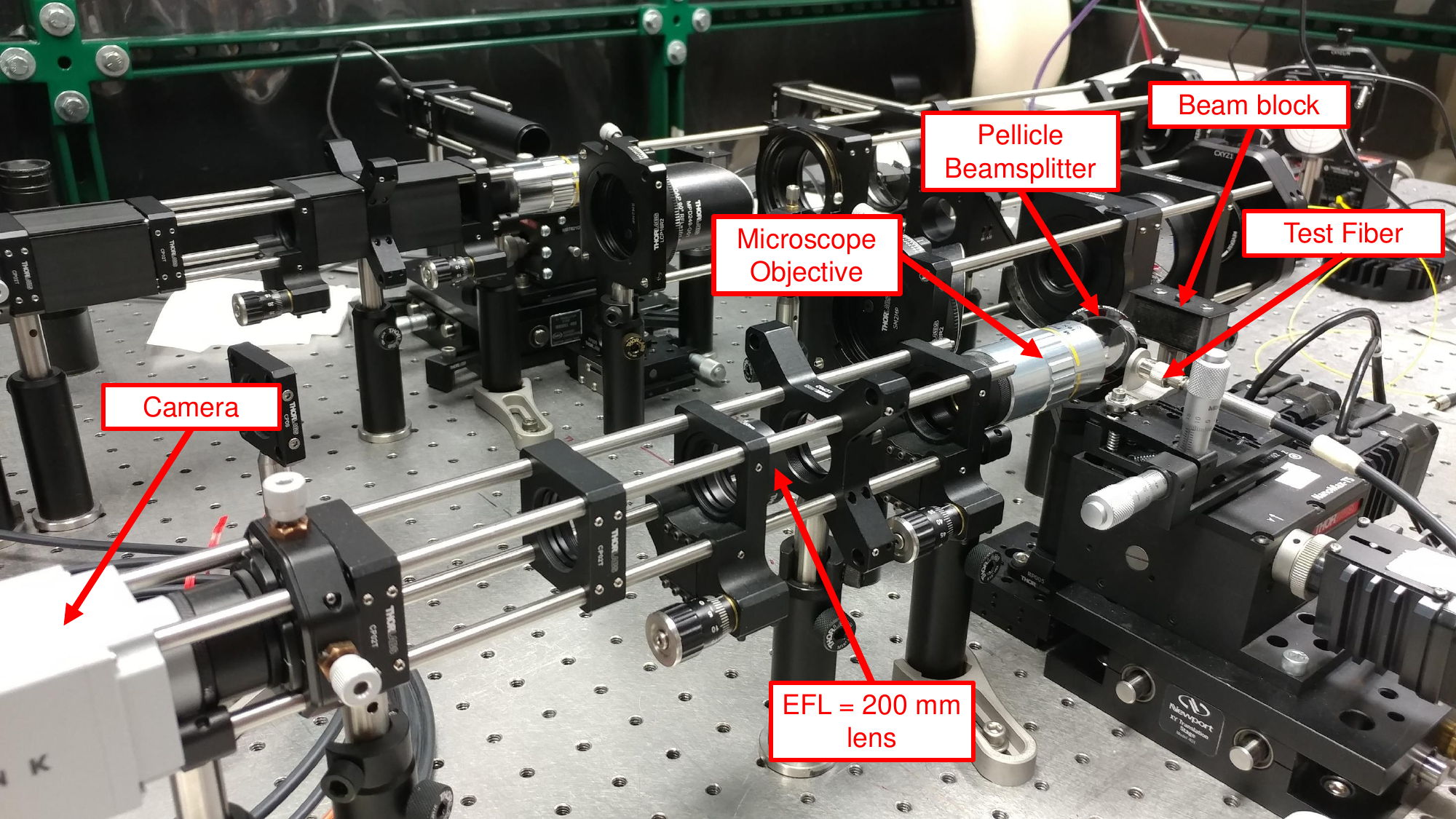}
    \caption{Injection imaging arm, with components labeled}
    \label{fig:FCS_injectimg}
\end{figure}

The injection imaging arm, shown in Figures \ref{fig:FCS_zmx_injectimg} and \ref{fig:FCS_injectimg}, images the test fiber input face to ensure that the input light is correctly fed into the test fiber. It looks back at the fiber input face via the pellicle beamsplitter. This arm also enables experiments of scrambling gain\cite{Avila2006,Halverson2015} (not to be confused with mode scrambling) to be done. The pellicle beamsplitter is mounted onto a magnetic kinematic base (Thorlabs model KB1X1) for easy removal. This apparatus is then mounted on a rotation stage (Thorlabs model RP005), which is mounted onto a XY translation stage for alignment purposes. The reflected light from the fiber input face is reflected by the pellicle beamsplitter into the injection imaging arm. Compared to the pre-injection arm which uses a 60 mm cage system, the injection imaging arm uses a 30 mm cage system. When light enters the injection imaging arm, it is first collimated by a Mitutoyo MY10X-803 infinite conjugate microscope objective lens (EFL $= 20$ mm). The objective lens is held in place by a Thorlabs SM1A27 adapter and a Z-axis translation mount (Thorlabs model SM1ZA) with a micrometer, for alignment purposes.

The collimated beam is then focused by a Ø$1''$ 200 mm EFL achromatic doublet lens (Thorlabs model AC254-200-A-ML) onto a detector. This lens is also held in place by a Z-axis translation mount. The detector is a Pixelink PL-B781U CMOS camera ($2208 \times 3000$ pixels, 3.5 \textmu m pixel pitch), and is held in place by a XY translation mount (Thorlabs model CXY1A). Note that the objective lens and 200 mm EFL lens together create a microscope with a magnification of 10. On the detector, the size of the image of the fiber input face is 10 times that of the fiber input face.

The objective lens and camera cause sag in the cage system due to their mass, and hence are both supported by additional mounting brackets (Thorlabs model CP02B). In addition, note that along the cage rail, between the 200 mm EFL lens and the objective lens, is a Thorlabs LCP02 cage plate adapter which converts between the 60 mm cage system and the 30 mm cage system. This adapter is used to aid in alignment.


\subsection{Near-Field Arm}\label{sec:near_field_arm}

\begin{figure}[H]
    \centering
    \includegraphics[width=0.75\textwidth]{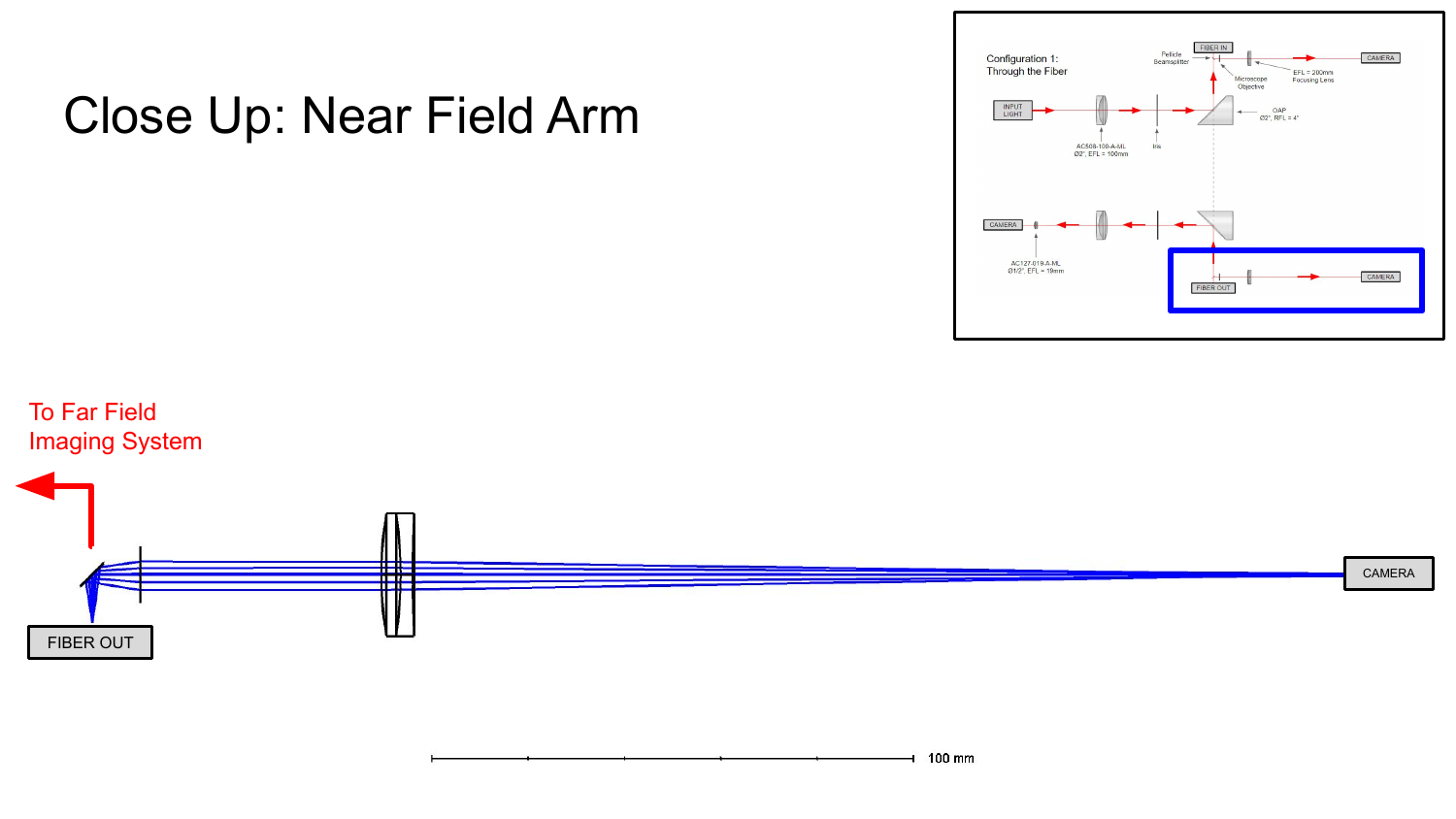}
    \caption{Zemax OpticStudio model of the near-field arm. The inset at the top right shows the location of these components in Figure \ref{fig:FCS_config1}.}
    \label{fig:FCS_zmx_near_field}
\end{figure}

The light which is injected into the test fiber travels through the fiber and exits the fiber on the other side of the FCS, which is nearly a mirror copy of the injection side. A pellicle beamsplitter (same model as the one on the injection side) splits the outputted cone of light from the test fiber into two directions. Some of the light is reflected into the near-field arm, which images the near-field of the test fiber (output face of the test fiber). The near-field arm, shown in Figure \ref{fig:FCS_zmx_near_field}, is nearly identical to the injection arm, except that a different detector is used. Instead, a Matrix Vision mvBlueCOUGAR-X102kG CMOS camera ($1600 \times 1104$ pixels, 9 \textmu m pixel pitch) is used. In addition, cage system covers (Thorlabs model C30L24) were used for the near-field arm in order to reduce the effects of stray light. This near-field arm takes the images which we analyze for modal noise using our image analysis procedure in \mbox{Section \ref{sec:analysis}}.

\subsection{Far-Field Arm}\label{sec:far_field_arm}

\begin{figure}[b]
    \centering
    \includegraphics[width=0.8\textwidth]{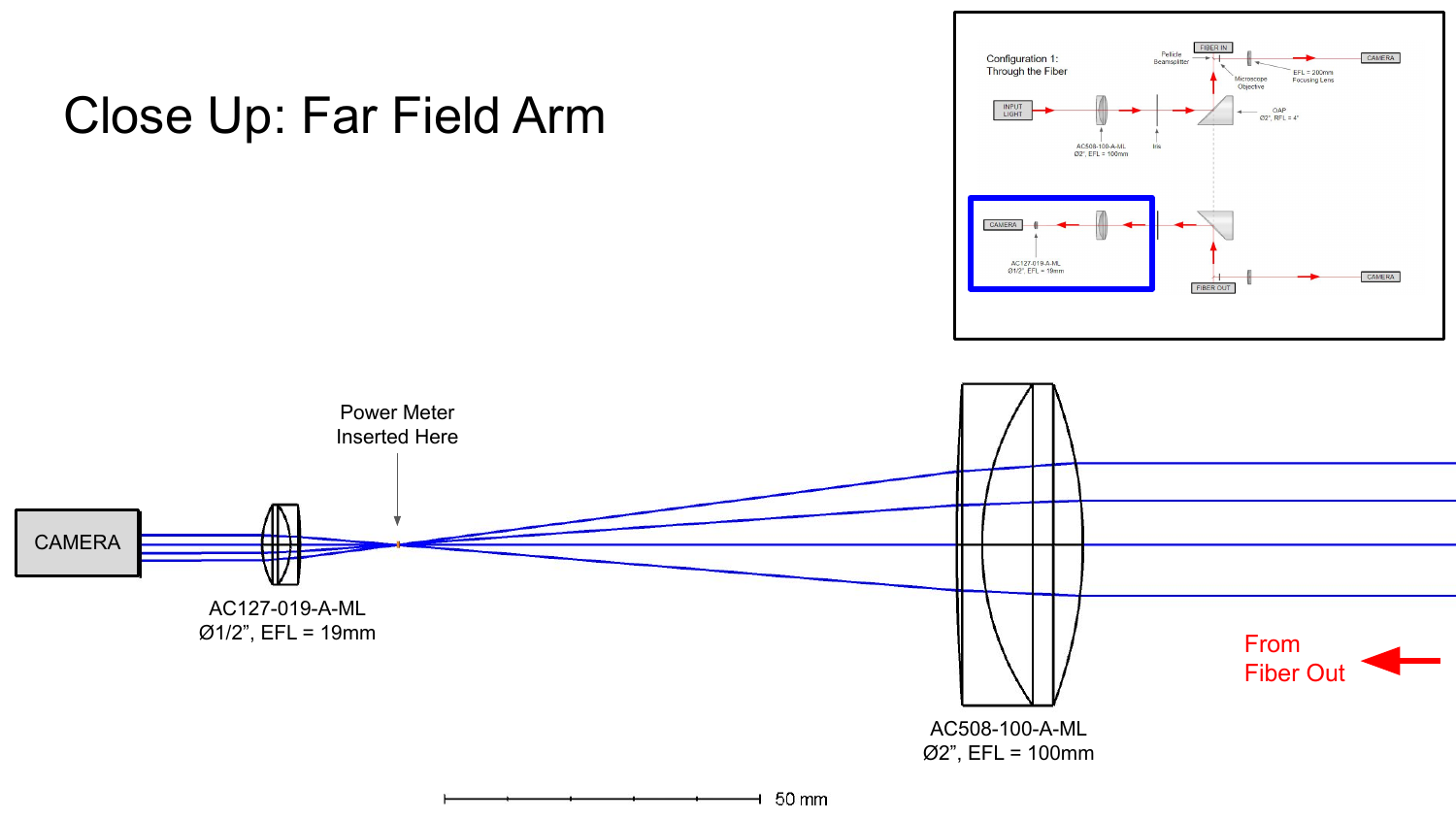}
    \caption{Zemax OpticStudio model of part of the far-field arm, from the 100 mm EFL lens to the Matrix Vision camera. The inset at the top right shows the location of these components in Figure \ref{fig:FCS_config1}.}
    \label{fig:FCS_zmx_far_field}
\end{figure}

\begin{figure}[t]
    \centering
    \includegraphics[width=0.75\textwidth]{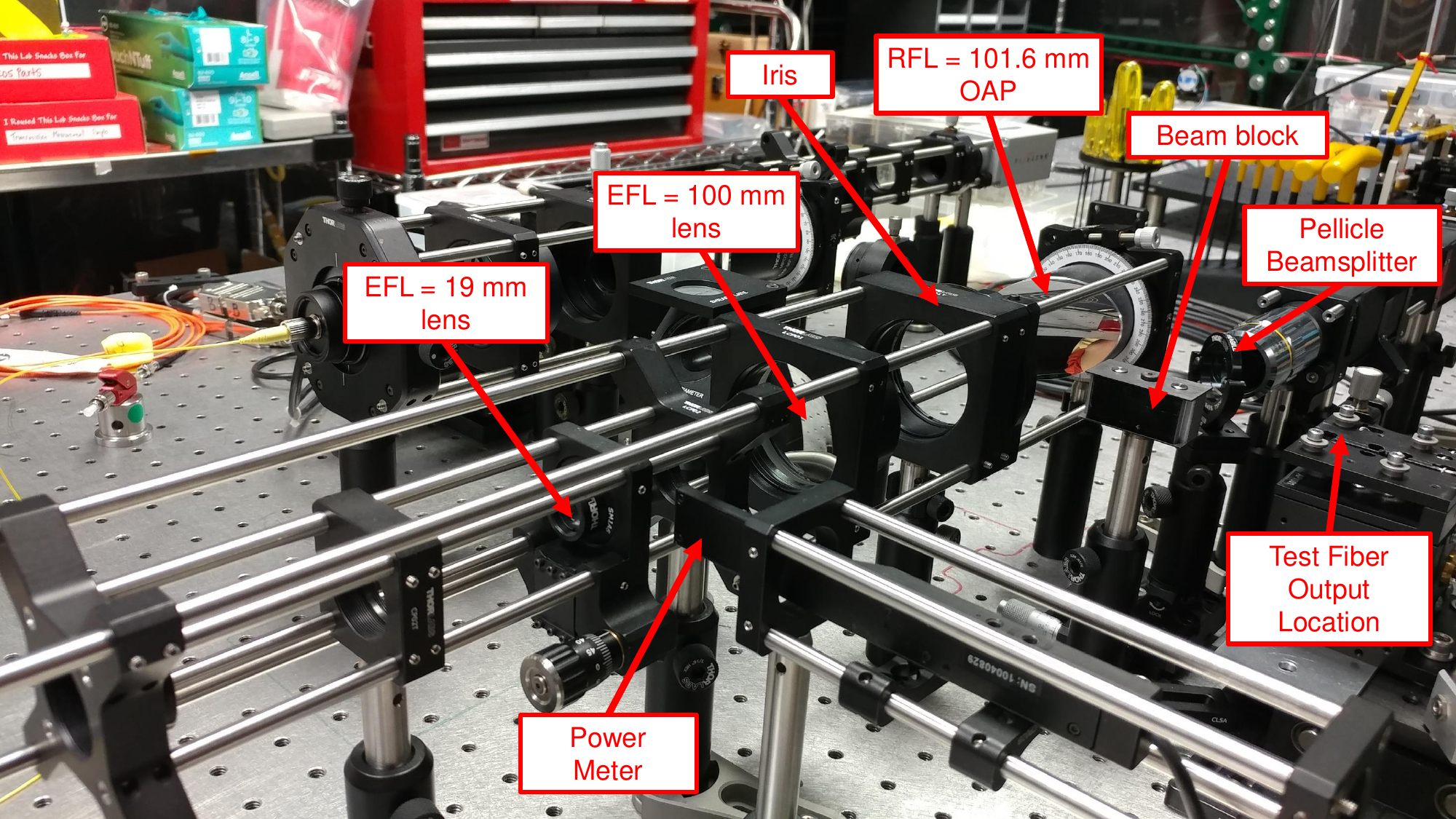}
    \caption{Far-field arm, with components labeled}
    \label{fig:FCS_farfield}
\end{figure}

The rest of the light which is not reflected by the pellicle beamsplitter into the near-field arm instead passes through the beamsplitter towards the far-field arm, shown in Figures \ref{fig:FCS_zmx_far_field} and \ref{fig:FCS_farfield}. The far-field arm is the same as the pre-injection arm, up until a 100 mm EFL lens. The output cone of light from the fiber is collimated by a Ø$2''$ $90^\circ$ OAP with 101.6 mm RFL, passes through an iris diaphragm, and then is focused by a Ø$2''$ 100 mm EFL achromatic doublet lens. At the focus is a power meter, which can slide in and out of the cage system using some rails.

If the power meter is not in place, then the focused beam expands and is recollimated by a Ø$0.5''$ 19 mm EFL achromatic doublet lens. Here, the cage system changes from the 60 mm cage system, down to the 30 mm cage system using a Thorlabs LCP02 cage plate adapter. The 19 mm EFL lens is held in place by Thorlabs SM1A6 adapter and a Z-axis translation mount (Thorlabs model SM1ZA) with a micrometer, for alignment purposes. After the 19 mm EFL lens, the 30 mm cage system changes back to the 60 mm cage system. The collimated beam is imaged by a Matrix Vision mvBlueCOUGAR-X102kG CMOS camera, which images the far-field of the fiber. The camera is held in place by a XYZ translation mount (Thorlabs model CXYZ1). The 100 mm EFL lens and 19 mm EFL lens together create a $\sim$5$\times$ beam reducer so that the far-field image could fit on the detector.

\section{Mode Scrambler Design}\label{sec:MS_design}

\subsection{Mechanical Design}

In this section, we discuss the design of our mode scrambler. The mechanical design of our mode scrambler is a four-bar linkage crank-rocker\cite{Uicker}. An illustration of this crank-rocker design is shown in Figure \ref{fig:four-bar} below. In this figure, points $M$ and $Q$, at the triangle, are fixed. Bar $MQ$ is fixed and does not move. A stepper motor is attached to point $M$ and rotates bar $AM$ continuously. The fiber is attached to bar $BQ$, which rotates about point $Q$ and agitates the fiber. For a crank-rocker, $a+b<c+d$.
\begin{figure}[H]
    \centering
    \includegraphics[width=0.47\textwidth]{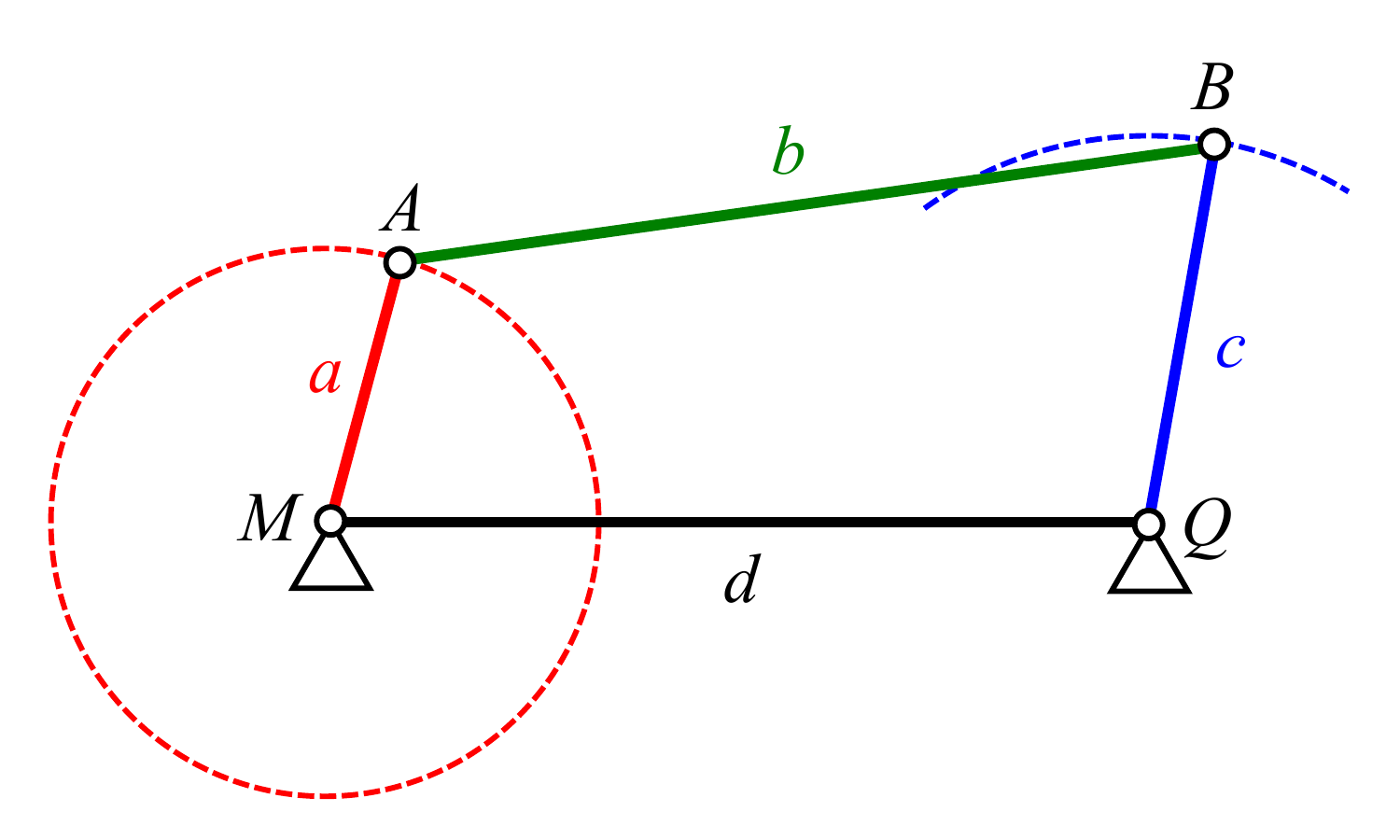}
    \caption{Four-bar linkage crank-rocker illustration}
    \label{fig:four-bar}
\end{figure}

\begin{figure}[H]
    \centering
    \includegraphics[width=0.7\textwidth]{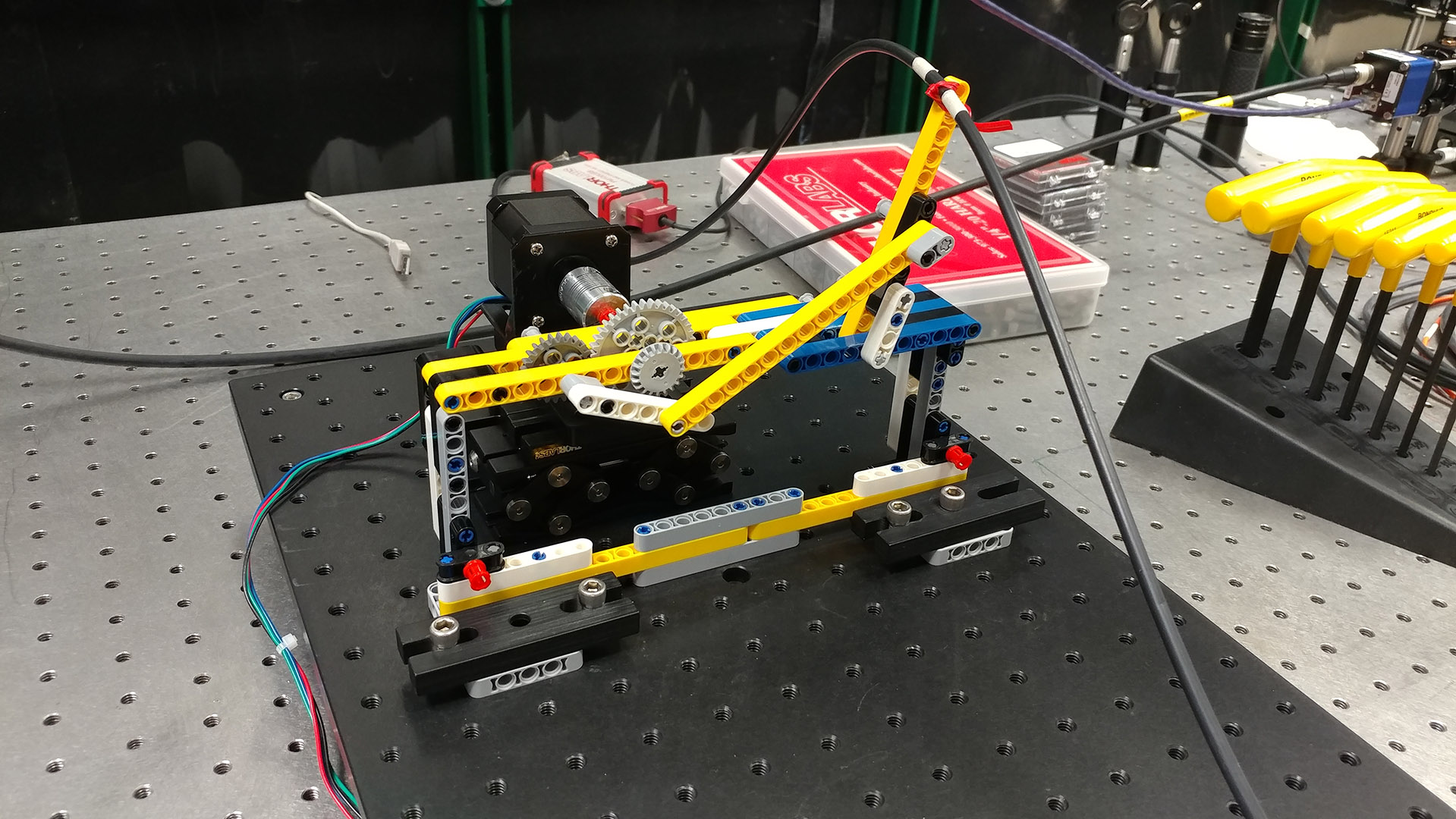}
    \caption{Prototype mode scrambler with a four-bar linkage crank-rocker design, built using LEGO Technic elements. Compare with Figure \ref{fig:four-bar}. The test fiber is the black cable, and is tied down to bar $BQ$. Note that here, the motor shaft is not directly attached to point $M$, but instead a gear system is used with a 40:24 gear ratio.}
    \label{fig:four-bar_lego}
\end{figure}

In our design, the lengths of the bars satisfied $b=d = 3a$ and $a = \frac{5}{7}c$. An example set of measurements is $b=d=15$, $a=5$, and $c=7$. When this design was constructed, bar $BQ$ was actually made longer, and extended beyond point $B$, allowing for the fiber to take on more positions along the bar. The mode scrambler was prototyped using LEGO® Technic™ elements, shown in Figure \ref{fig:four-bar_lego}.

\begin{figure}[H]
    \centering
    \includegraphics[width=0.7\textwidth]{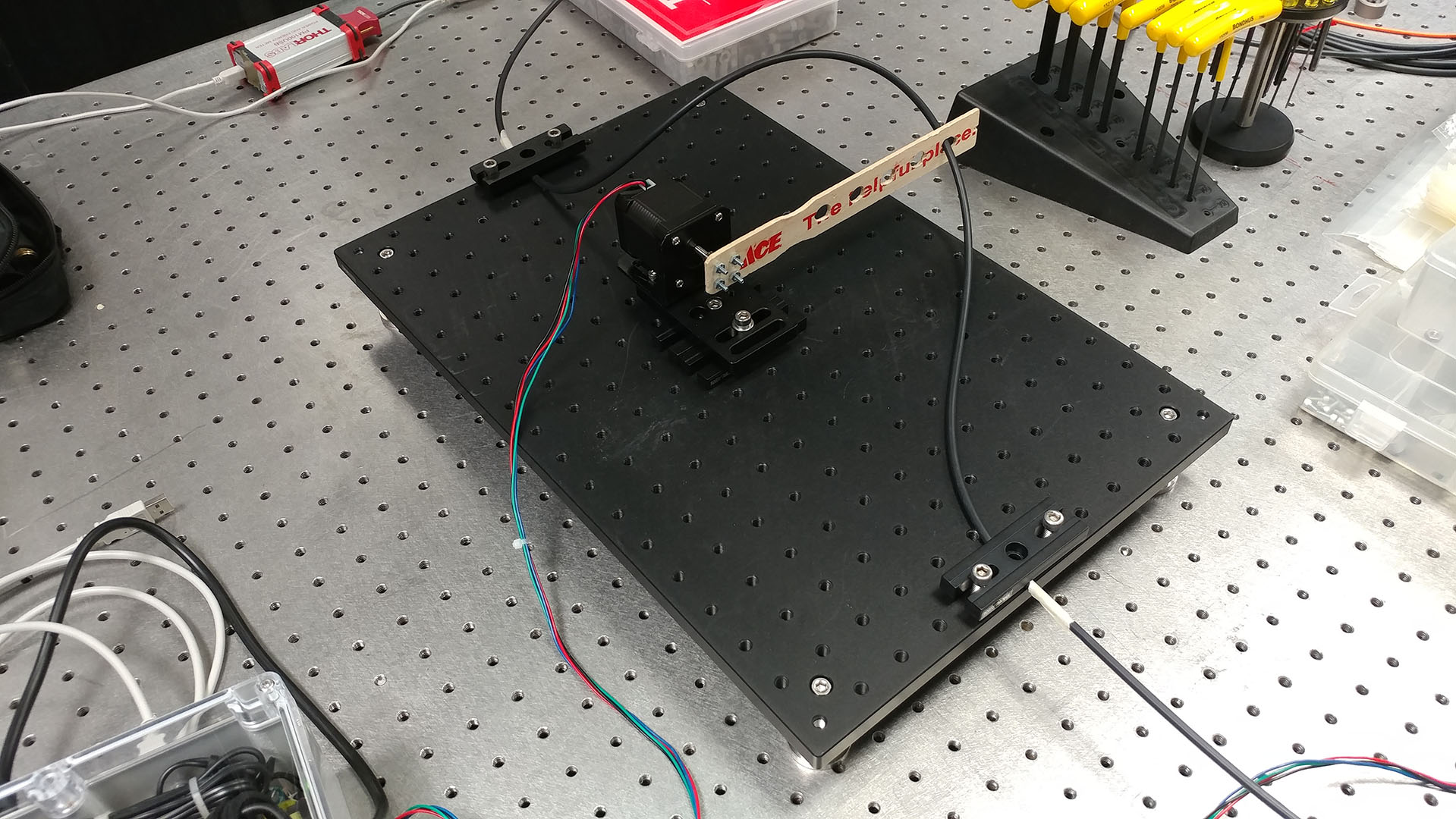}
    \caption{First prototype mode scrambler with a single rotating arm. The test fiber is the black cable.}
    \label{fig:single_arm}
\end{figure}

We initially started with an even simpler mechanical design, shown in Figure \ref{fig:single_arm}, which consists of a single rotating arm. We attached a wooden stick with a series of holes spaced $1''$ apart, at distances from $3''$ to $7''$ away from the motor shaft, for a total of five holes. A test fiber passes through the holes drilled into the arm, and when the arm is rotated, the fiber is agitated. When the mode scrambler is turned on, the arm rotates back and forth. In this design, the fiber could be passed through any one of the five holes, or all of the holes at once. The distance between the motor shaft and the fiber is a parameter that was varied. The four-bar linkage crank-rocker design in Figure \ref{fig:four-bar_lego} is an extension and improvement of this simpler design and has similar motion. The motor in the crank-rocker design continuously rotates in one direction, unlike in the single rotating arm design, which has to change direction of rotation and hence is more susceptible to slipping.

\subsection{Electrical and Hardware Design}\label{sec:MS_design_elec}

\begin{figure}[H]
    \centering
    \includegraphics[width=0.80\textwidth]{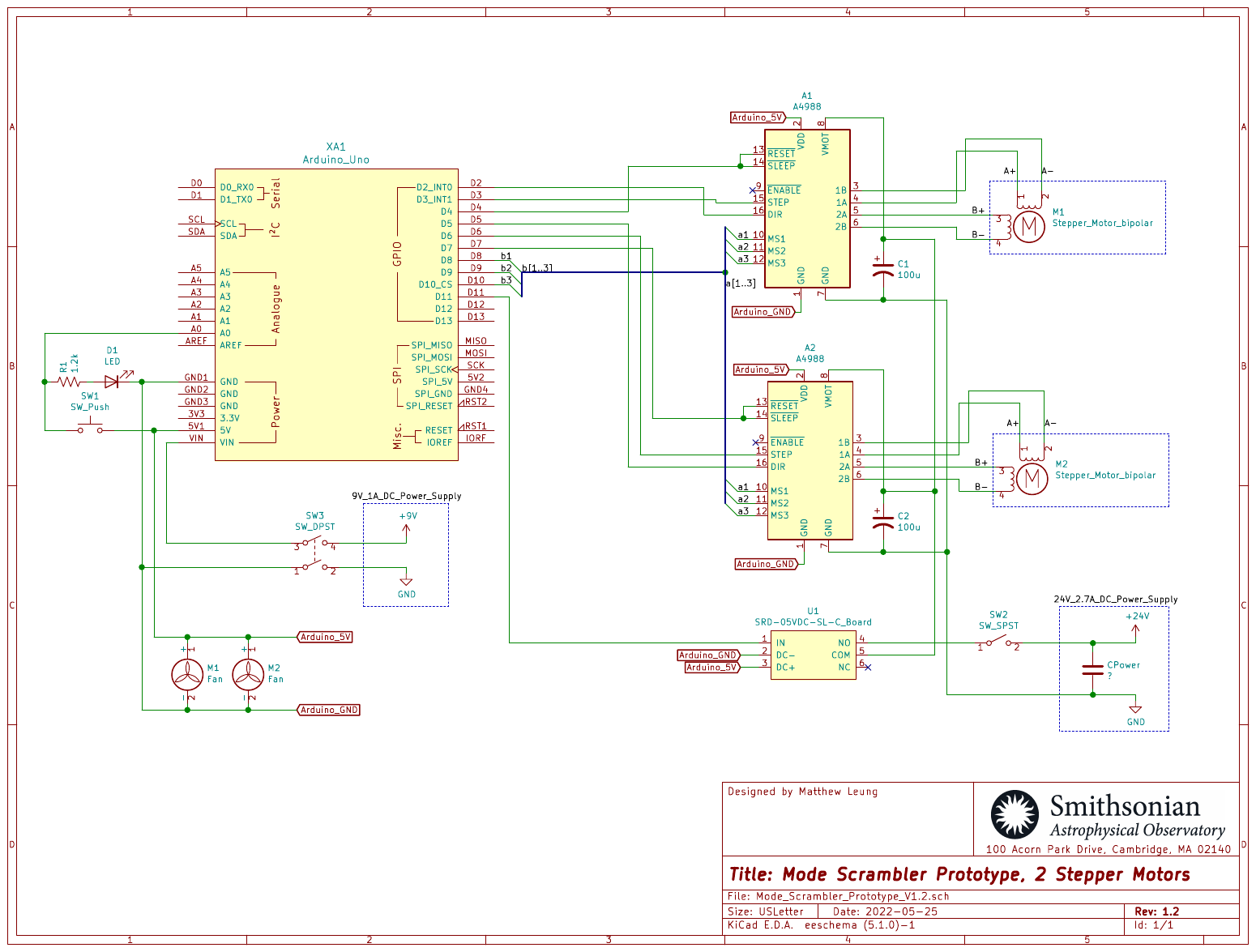}
    \caption{Schematic of the electrical and hardware design for the mode scrambler prototype. The design here allows for two stepper motors to be connected.}
    \label{fig:elec_schematic}
\end{figure}

\begin{figure}[H]
    \centering
    \includegraphics[width=0.6\textwidth]{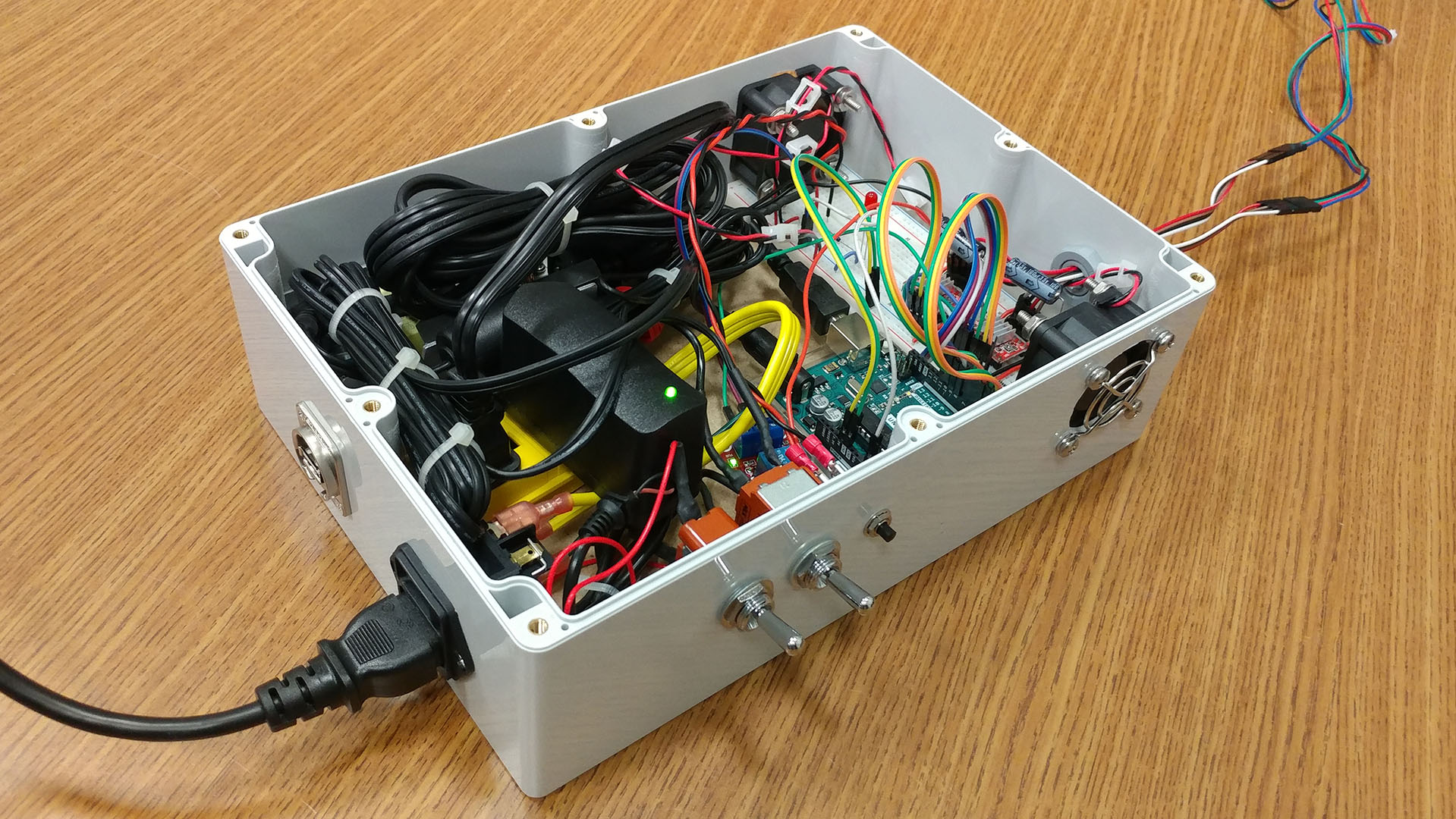}
    \caption{Electronics box housing the components in Figure \ref{fig:elec_schematic}.}
    \label{fig:elec_box}
\end{figure}

A schematic of the electrical and hardware design for the mode scrambler prototype is shown in Figure \ref{fig:elec_schematic}. We used a NEMA 17 bipolar stepper motor with 59 $\textrm{N}\cdot\textrm{cm}$ torque to drive bar $AM$. The stepper motor (M1 in schematic) is controlled by an Arduino UNO R3 (XA1 in schematic), using an A4988 stepper motor driver (A1 in schematic). The motors are powered by a 24 V 2.7 A DC power supply, shown on the bottom right of the schematic. A relay (U1 in schematic) shuts off the power to the motors in the event that the Arduino loses power. Two fans were included to cool the stepper motor drivers. The electrical and hardware components were housed in an electronics box, shown in Figure \ref{fig:elec_box}.

\newpage
\section{Image Analysis}\label{sec:analysis}

In this section, we discuss our image analysis procedure for near-field images of an optical fiber. The near-field arm uses a Matrix Vision CMOS camera which takes 12-bit images. During each experiment, in addition to taking images of the near-field (science frames), we took dark frames, bias frames, and flat field frame. After obtaining a corrected image (dark and bias subtracted, and flat field corrected), the main goal is to identify the boundary of the fiber face in the corrected image. Then some metric for scrambling can be computed from the pixels inside the fiber face boundary.

\subsection{Dark, Bias, and Flat Field Correction}\label{sec:MS_analysis_dbf}

While conducting each experiment, dark frames, bias frames, and flat field frames were taken in addition to the science frames. The dark frames had the same exposure time as the science frames and were taken when a lens cap was covering the camera. The bias frames were taken with the shortest exposure time possible for the camera, and were taken when a lens cap was covering the camera. The flat field frames were taken with an evenly-illuminated diffuse light source illuminated into the near-field arm. The diffuse light source consisted of two Thorlabs ground glass diffusers (Thorlabs model DG10-120) in front of a 60 mm EFL lens which collimated a white light LED.

Let SCI, DARK, BIAS, and FLAT be the science, dark, bias, and flat frames respectively. Let $t_\textrm{SCI}$, $t_\textrm{DARK}$, and $t_\textrm{FLAT}$ be the exposure times of the science, dark, and flat frames respectively. Then the corrected image can be found by\cite{BUflat}:
\begin{equation}
    \textrm{CORRECTED} = \textrm{FLATCORRECTED}_\textrm{avg} \times
    \left[ \frac{(\textrm{SCI} - \textrm{BIAS}) - \frac{t_\textrm{SCI}}{t_\textrm{DARK}} (\textrm{DARK} - \textrm{BIAS})}{(\textrm{FLAT} - \textrm{BIAS}) - \frac{t_\textrm{FLAT}}{t_\textrm{DARK}} (\textrm{DARK} - \textrm{BIAS})} \right]
\end{equation}
where:
\begin{equation}
    \textrm{FLATCORRECTED}_\textrm{avg} = \overline{\left[(\textrm{FLAT} - \textrm{BIAS}) - \frac{t_\textrm{FLAT}}{t_\textrm{DARK}} (\textrm{DARK} - \textrm{BIAS}) \right]}
\end{equation}
is the average value of the flat frame (the overline denotes an average), corrected for bias and dark effects.


\subsection{Fiber Face Boundary Identification}\label{sec:fiber_boundary}

After dark, bias, and flat field corrections, the next step is to identify the pixels in the corrected image which belong to the fiber face. After these pixels are identified, metrics can be computed for scrambling. The problem is as follows: given an image of a face of a fiber (near-field in this case), we seek to identify the boundary of the fiber. 

To do this, several different approaches were attempted. Initial approaches involved taking spatial derivatives of the image, using Sobel filters or Laplacian filters, in order to make the fiber boundary (edge) feature more apparent. A Canny edge detector\cite{Canny} was then applied to the filtered image (with the spatial derivative filters applied) in order to identify the boundary of the fiber face. Afterwards, the result of the Canny edge detector (which gives points representing edges) were converted into contours (lines joining the points). The longest contour was anticipated to represent the boundary of the fiber face, and an attempt was made to use Douglas-Peucker algorithm\cite{DOUGLAS1973} to further approximate the longest contour to reduce jaggedness in the boundary shape. However, the result of the Canny edge detector suffered from broken edges, which is a limitation commonly mentioned the literature \cite{Akinlar2016,Dhillon2022}. The resulting contours do not represent the fiber boundary well. Thresholding was also attempted on the Laplacian-filtered or Sobel-filtered image before applying the Canny edge detector, but the results were not too generalizable.

Instead of using more edge detection methods, another approach was attempted. It is important to note that the face of an optical fiber is a certain shape (e.g. circle, square, rectangle, octagon). The boundary of an optical fiber is not just some arbitrary edge. This is important information which points to computational geometry approaches instead for solving the problem of identifying the fiber face boundary. It was realized that the boundary of an optical fiber which has a convex shape (e.g., square and octagon) can be represented by a convex hull. However, some fibers may have more complicated shapes which are concave instead (e.g., a rectangular fiber which was tested in this project). The boundary of such a fiber would need to be represented by a ``concave hull''. The idea of a convex hull can be generalized to what is called an alpha shape \cite{Edelsbrunner1994}. A set of finite points can be obtained representing the entire fiber face, and the outer boundary of the alpha shape of this set of points would be the boundary of the fiber face. This approach turned out to be very effective, and is summarized by the following steps:
\begin{enumerate}
    \itemsep0em 
    \item Apply Canny edge detector to an 8-bit version of the corrected image, and obtain a binary image which represents the Canny edges.
    \item Take the nonzero points from the binary image, and find the alpha shape of these points.
    \item Find the outer boundary of the alpha shape.
    \item If desired, offset the alpha shape.
\end{enumerate}

In the approach involving alpha shapes, the following was done. Firstly, the 12-bit image was converted into an 8-bit image. Canny edge detector was applied to the 8-bit image using the OpenCV implementation\cite{opencv_library}, and a binary image was obtained which represents the Canny edges. Let $S$ be the set of points representing the position of the nonzero pixels in the binary image. The next step is to find the alpha shape of $S$, and there are several steps involved, following Edelsbrunner's algorithm \cite{Edelsbrunner1994,Fischer2000}. The alpha shape is parameterized by a parameter $\alpha$. If $\alpha = \infty$, then the alpha shape is the convex hull of $S$, and if $\alpha=0$, then the alpha shape is just individual points \cite{Gardiner2018}. This is how an alpha shape can be thought of as a generalized version of the convex hull. If $\alpha$ is not a big number, then the alpha shape can be thought of as a ``concave hull''.

\begin{figure}[b]
    \centering
    \begin{subfigure}{0.34\textwidth}
        \centering
        \includegraphics[width=\textwidth]{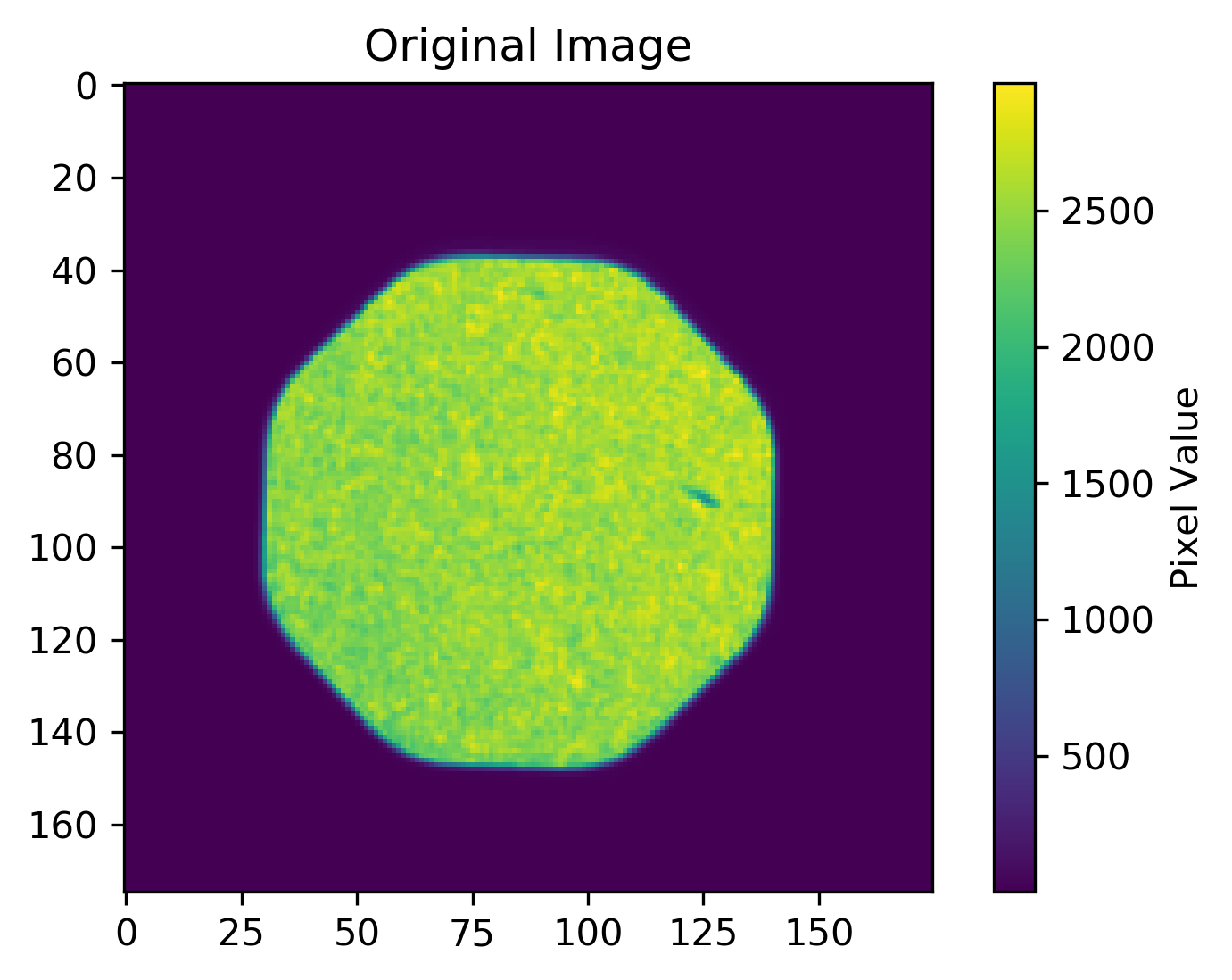}
        \caption{Original image}
        \label{fig:MS_analyze_AS_Oct_og}
    \end{subfigure}
    \begin{subfigure}{0.26\textwidth}
        \centering
        \includegraphics[width=\textwidth]{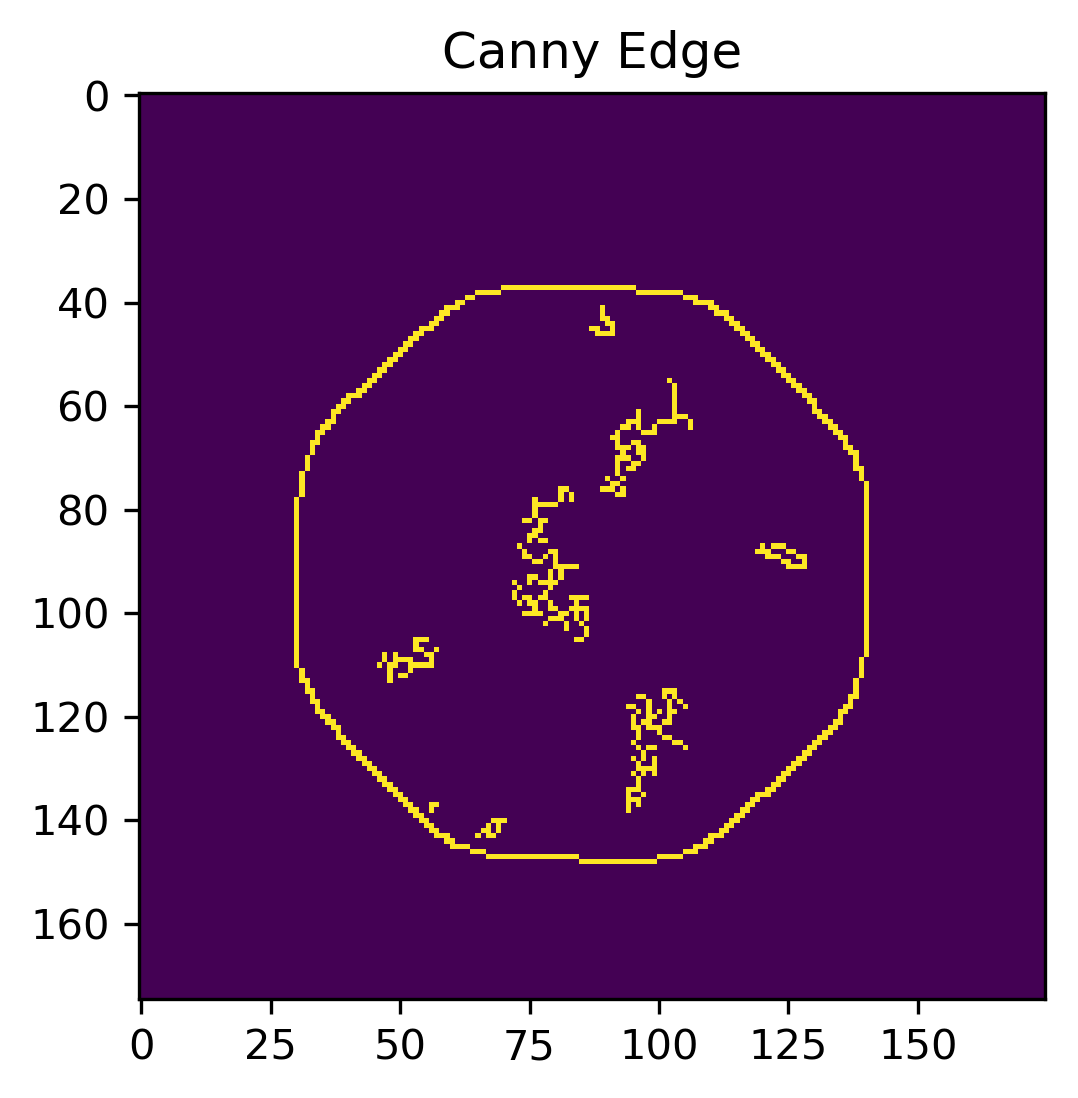}
        \caption{Set of points $S$}
    \end{subfigure}
    \begin{subfigure}{0.26\textwidth}
        \centering
        \includegraphics[width=\textwidth]{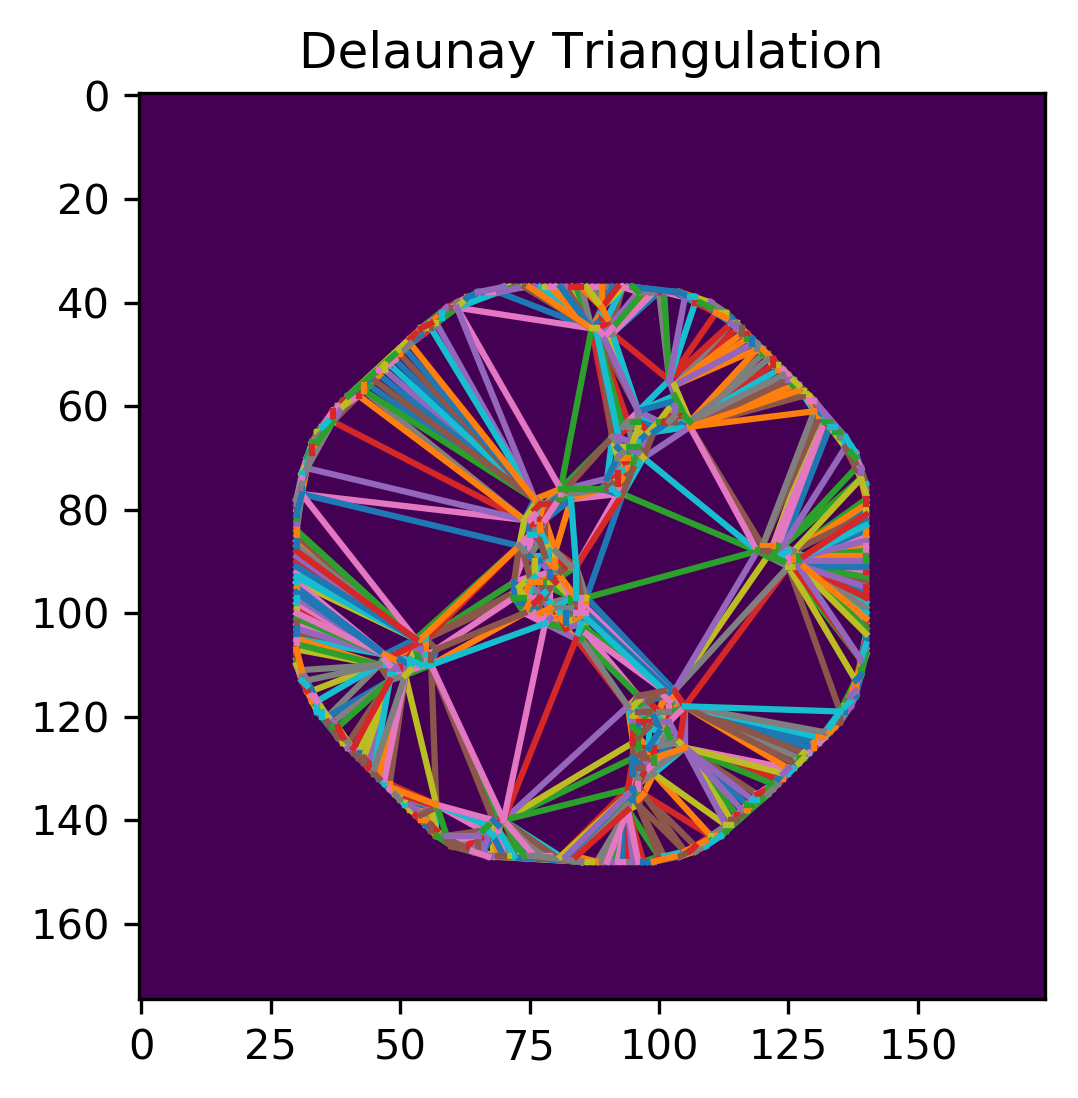}
        \caption{$\textrm{DT}(S)$}
    \end{subfigure}
    \begin{subfigure}{0.26\textwidth}
        \centering
        \includegraphics[width=\textwidth]{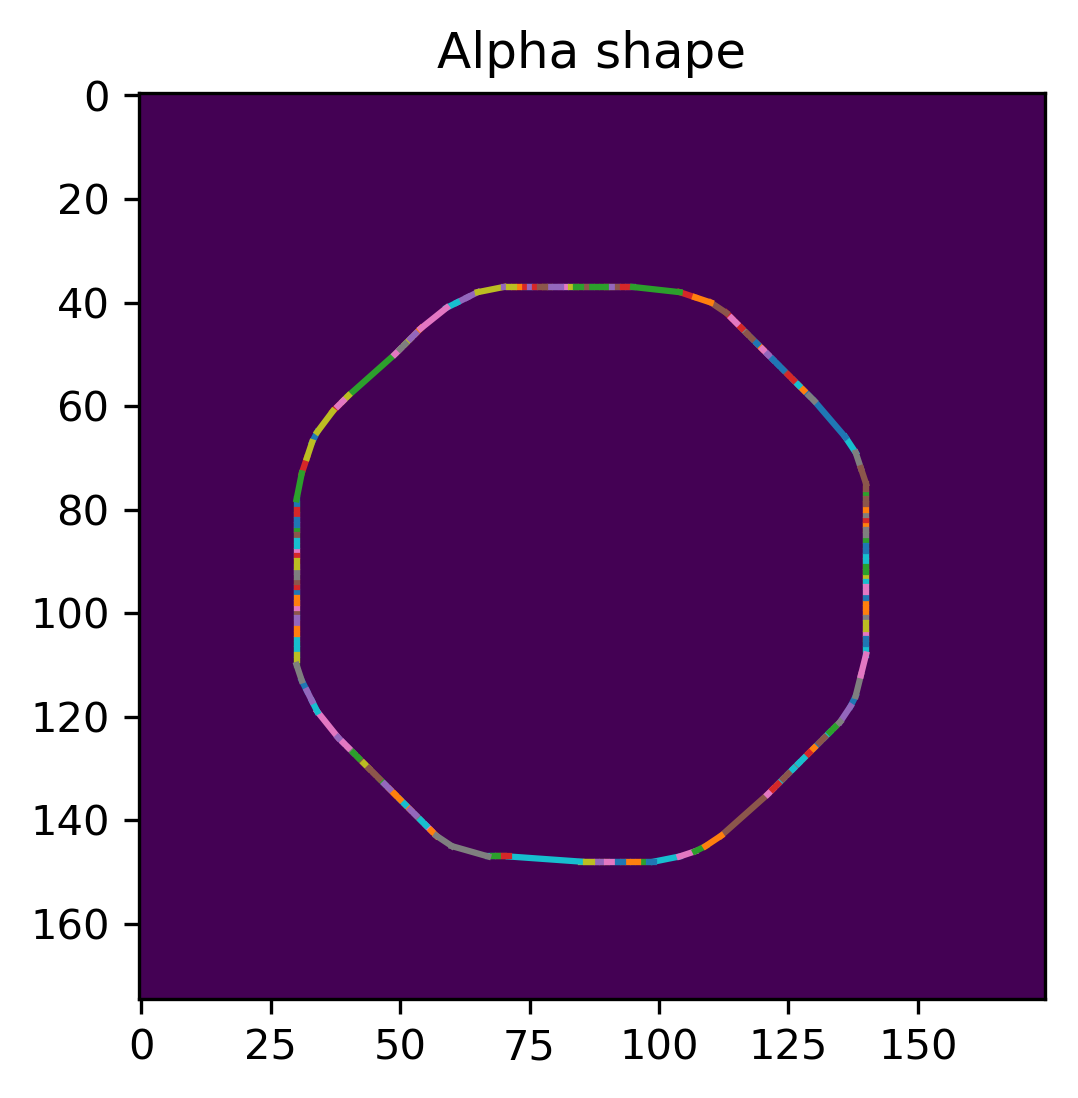}
        \caption{$\partial\mathcal{C}_\alpha(S)$}
    \end{subfigure}
    \begin{subfigure}{0.34\textwidth}
        \centering
        \includegraphics[width=\textwidth]{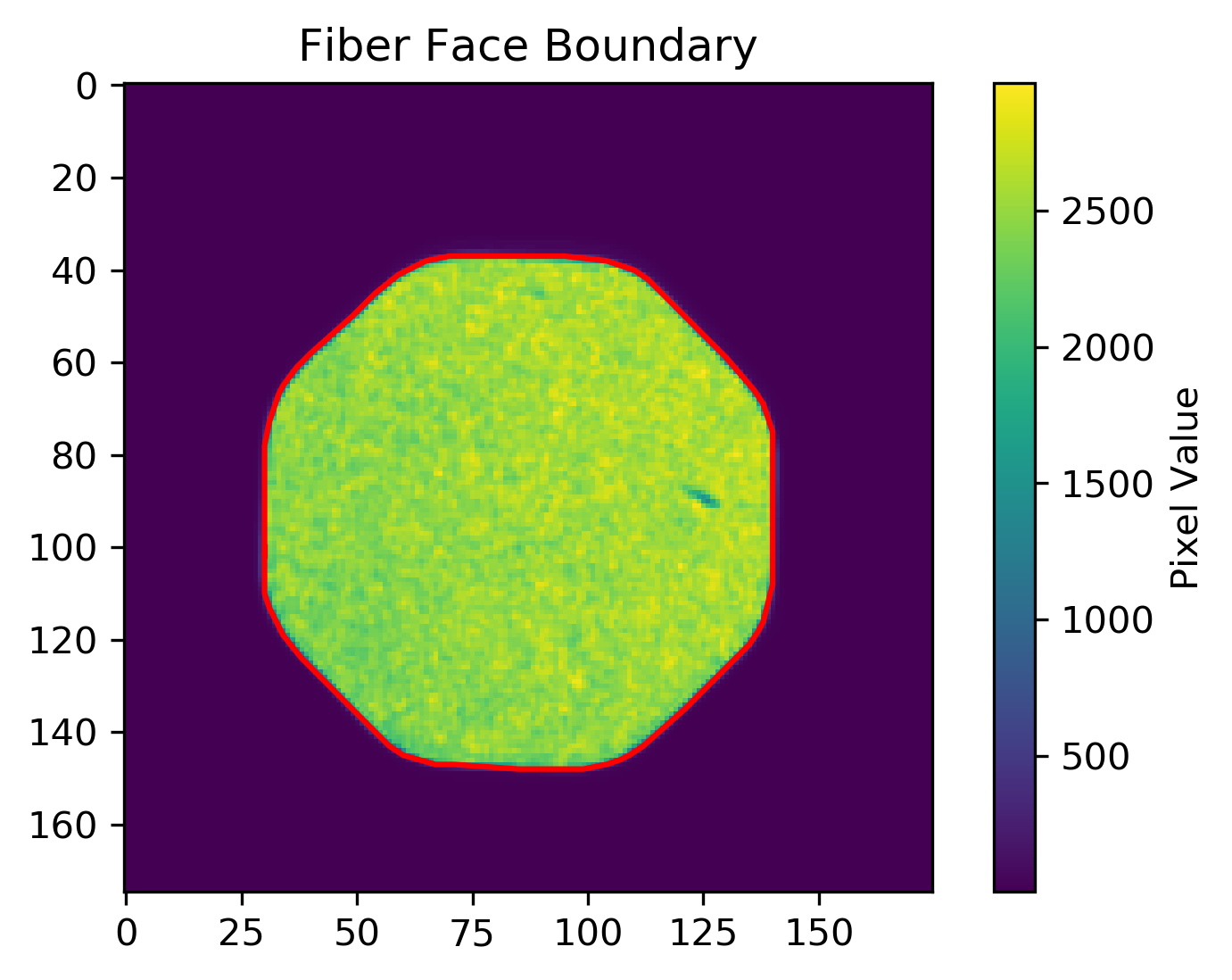}
        \caption{Fiber face boundary}
        \label{fig:MS_analyze_AS_Oct_boundary}
    \end{subfigure}
    \caption{Finding the fiber face boundary in a near-field image of a 100 \textmu m octagonal fiber (Figure \ref{fig:MS_analyze_AS_Oct_og}, taken when the mode scrambler was on, but with only little agitation), using the procedure described in Section \ref{sec:fiber_boundary}. Figure \ref{fig:MS_analyze_AS_Oct_boundary} shows the boundary in red. The axes represent pixel number.}
    \label{fig:MS_analyze_AS_Oct}
\end{figure}

\begin{figure}[th]
    \centering
    \begin{subfigure}{0.34\textwidth}
        \centering
        \includegraphics[width=\textwidth]{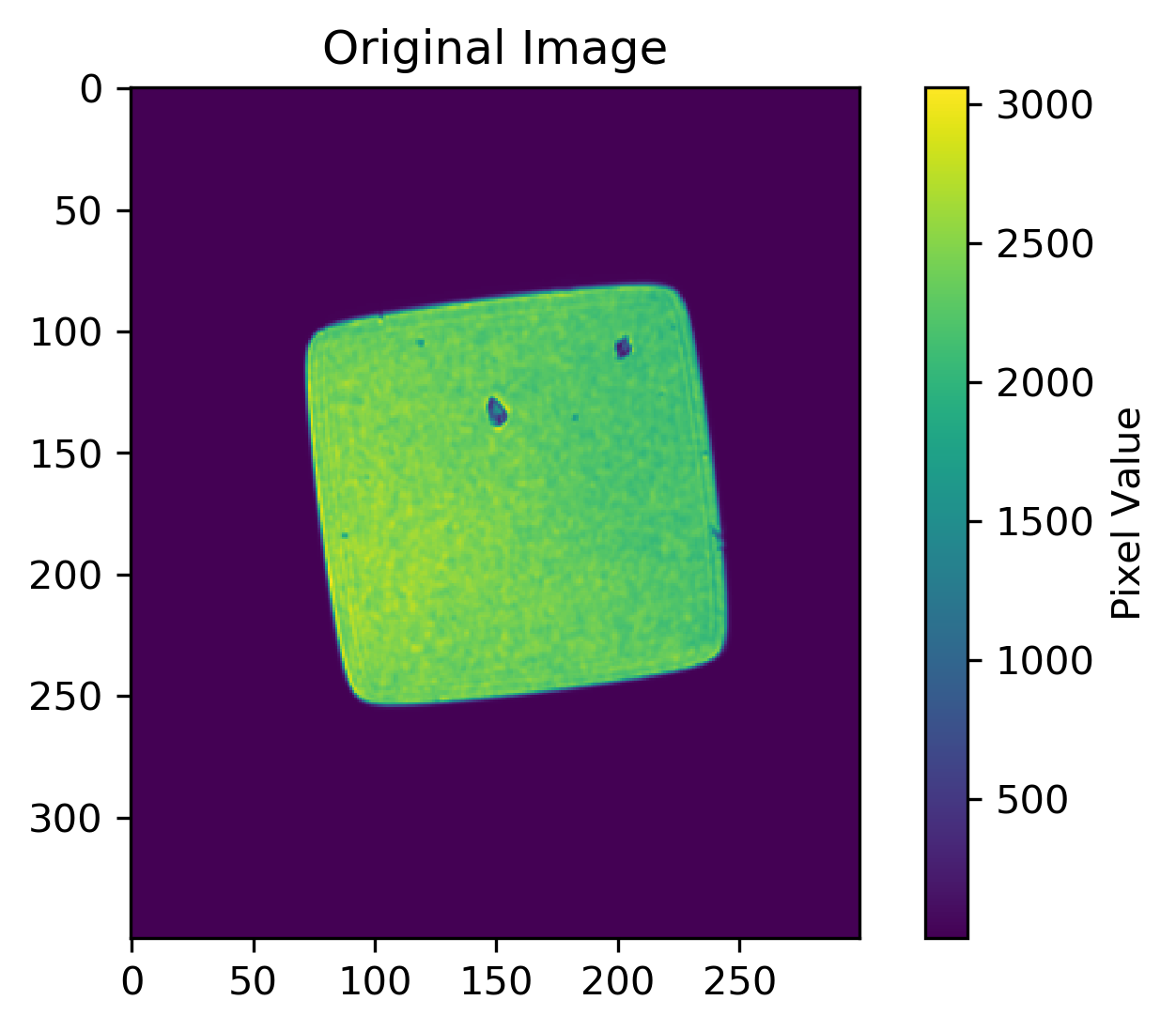}
        \caption{Original image}
        \label{fig:MS_analyze_AS_Sq_og}
    \end{subfigure}
    \begin{subfigure}{0.26\textwidth}
        \centering
        \includegraphics[width=\textwidth]{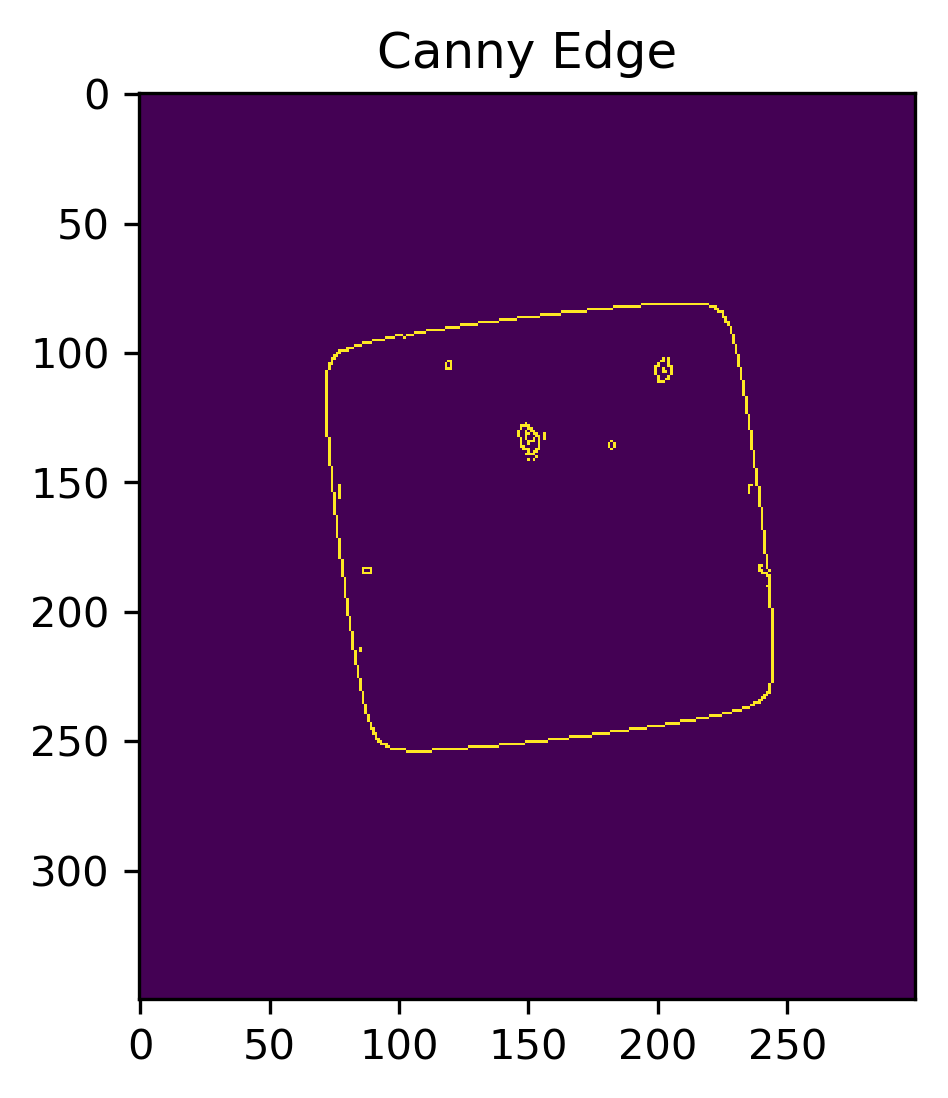}
        \caption{Set of points $S$}
    \end{subfigure}
    \begin{subfigure}{0.26\textwidth}
        \centering
        \includegraphics[width=\textwidth]{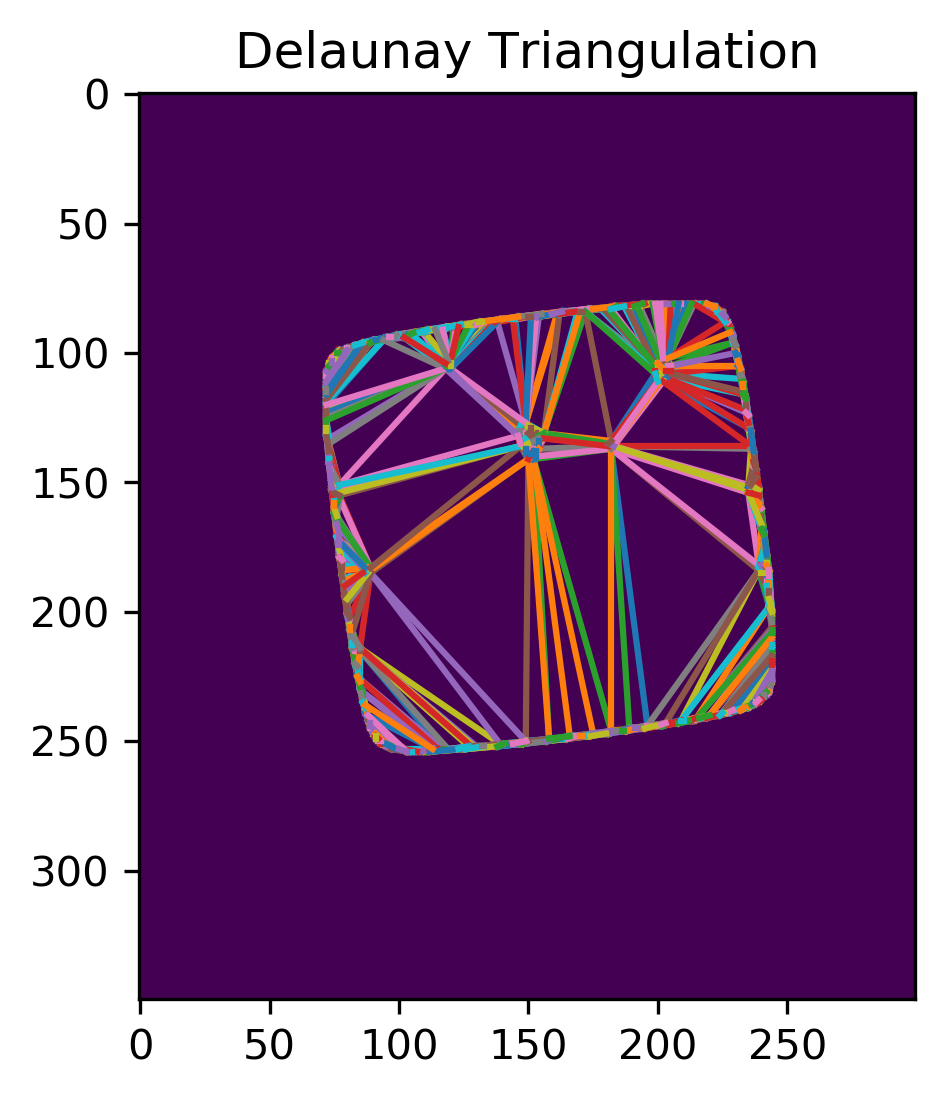}
        \caption{$\textrm{DT}(S)$}
    \end{subfigure}
    \begin{subfigure}{0.26\textwidth}
        \centering
        \includegraphics[width=\textwidth]{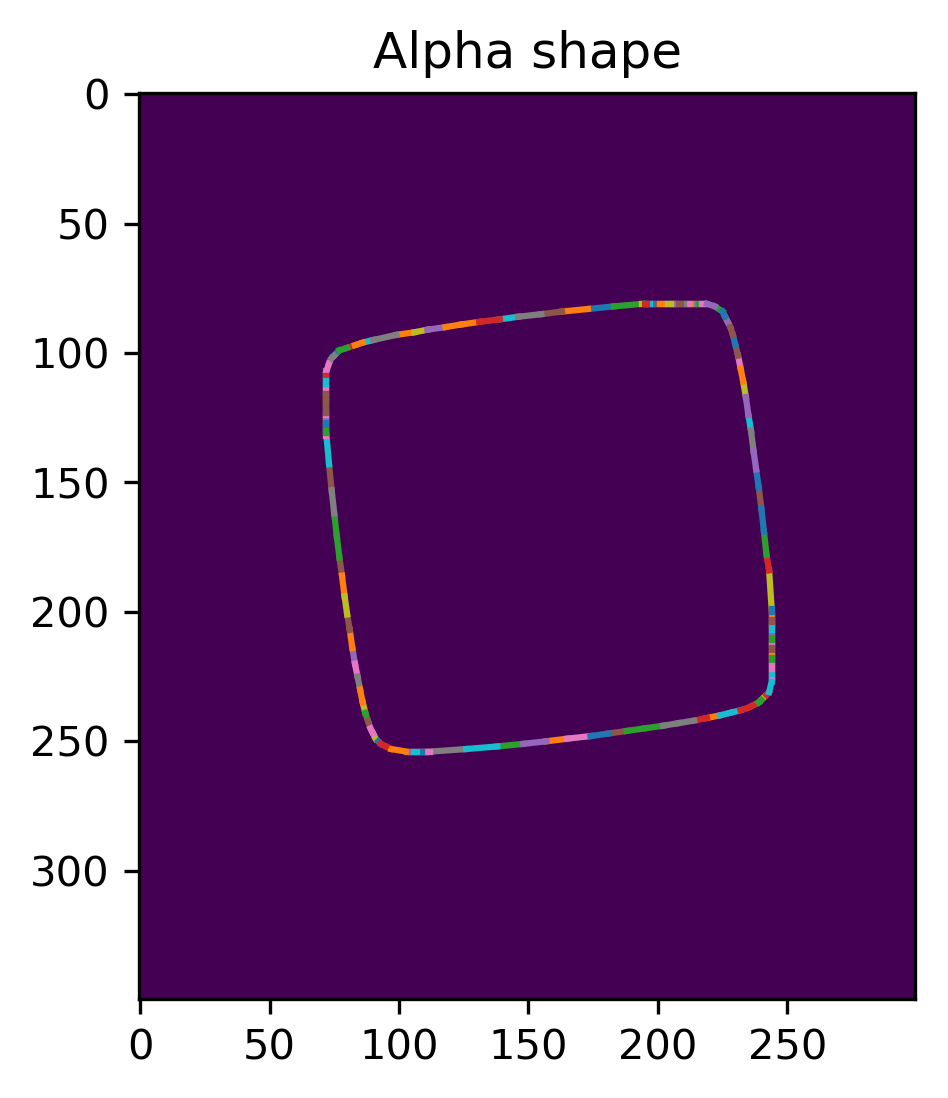}
        \caption{$\partial\mathcal{C}_\alpha(S)$}
    \end{subfigure}
    \begin{subfigure}{0.34\textwidth}
        \centering
        \includegraphics[width=\textwidth]{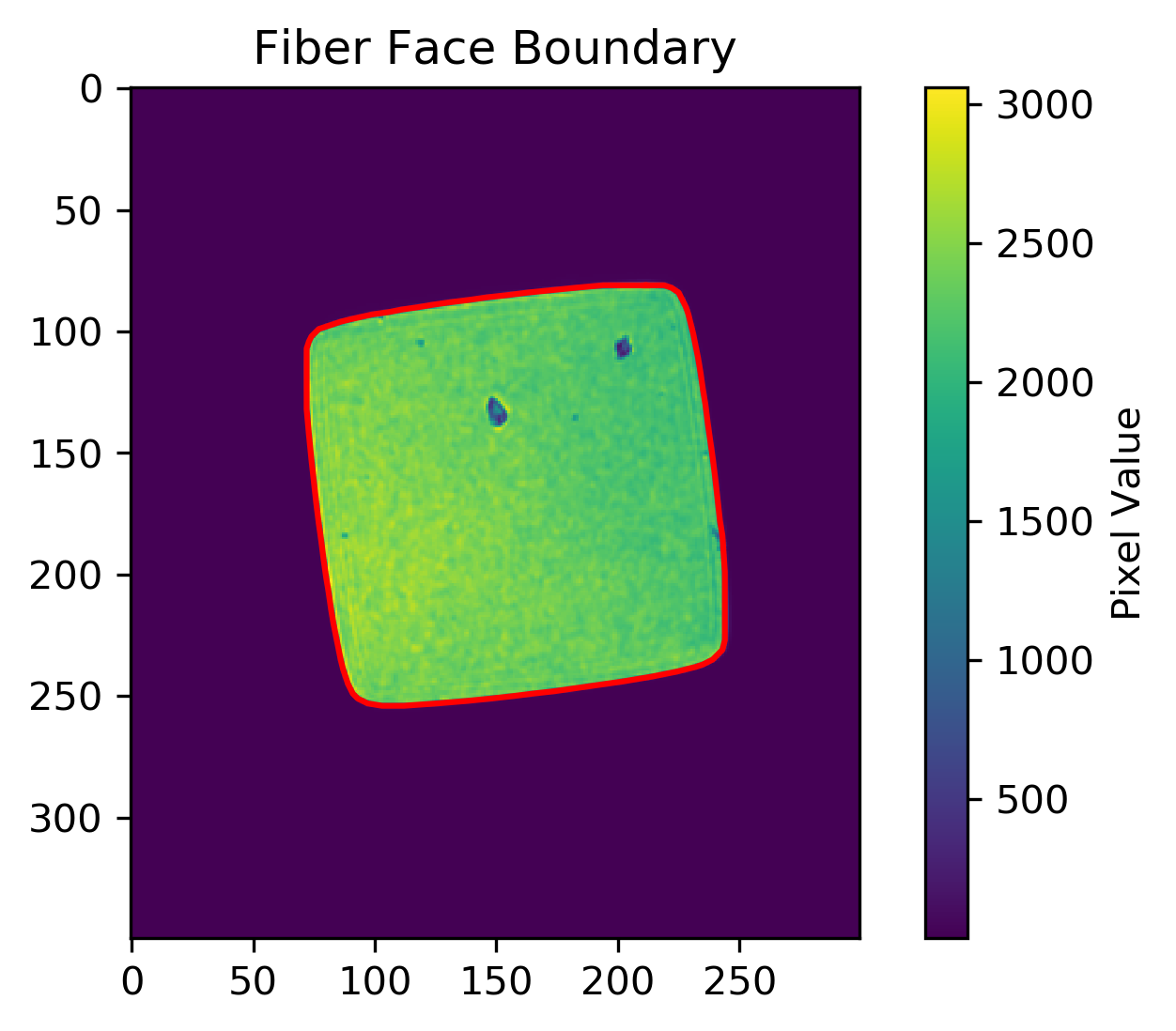}
        \caption{Fiber face boundary}
        \label{fig:MS_analyze_AS_Sq_boundary}
    \end{subfigure}
    \caption{Finding the fiber face boundary in a near-field image of a 150 \textmu m square fiber (Figure \ref{fig:MS_analyze_AS_Sq_og}, taken when the mode scrambler was on), using the procedure described in Section \ref{sec:fiber_boundary}. Figure \ref{fig:MS_analyze_AS_Sq_boundary} shows the boundary in red. The axes represent pixel number.}
    \label{fig:MS_analyze_AS_Sq}
\end{figure}

\begin{figure}[H]
    \centering
    \begin{subfigure}{0.3\textwidth}
        \centering
        \includegraphics[width=\textwidth]{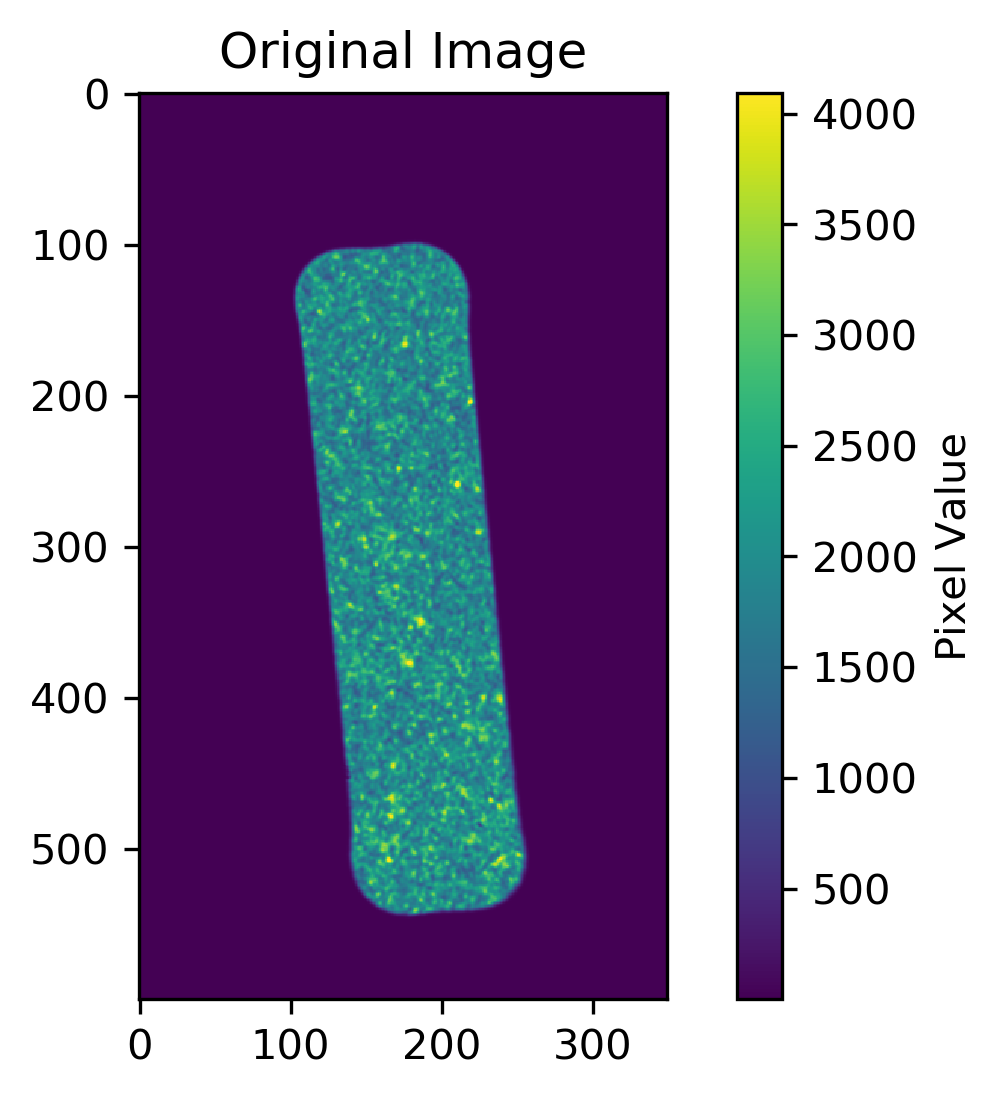}
        \caption{Original image}
        \label{fig:MS_analyze_AS_rect_og}
    \end{subfigure}
    \begin{subfigure}{0.21\textwidth}
        \centering
        \includegraphics[width=\textwidth]{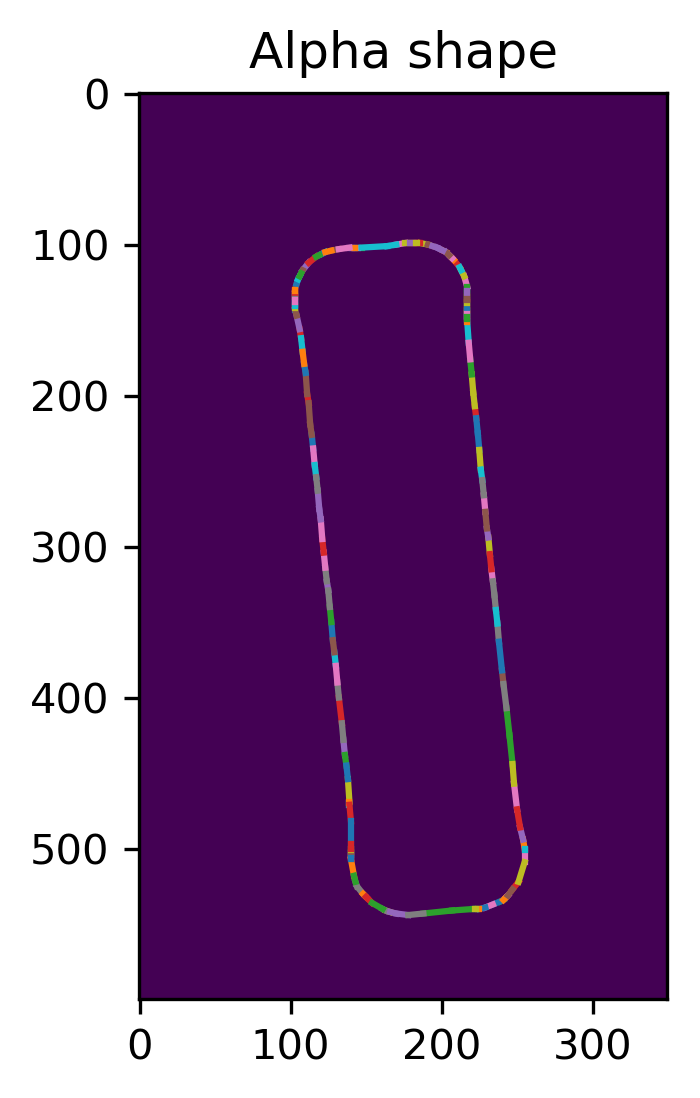}
        \caption{$\partial\mathcal{C}_\alpha(S)$}
    \end{subfigure}
    \begin{subfigure}{0.3\textwidth}
        \centering
        \includegraphics[width=\textwidth]{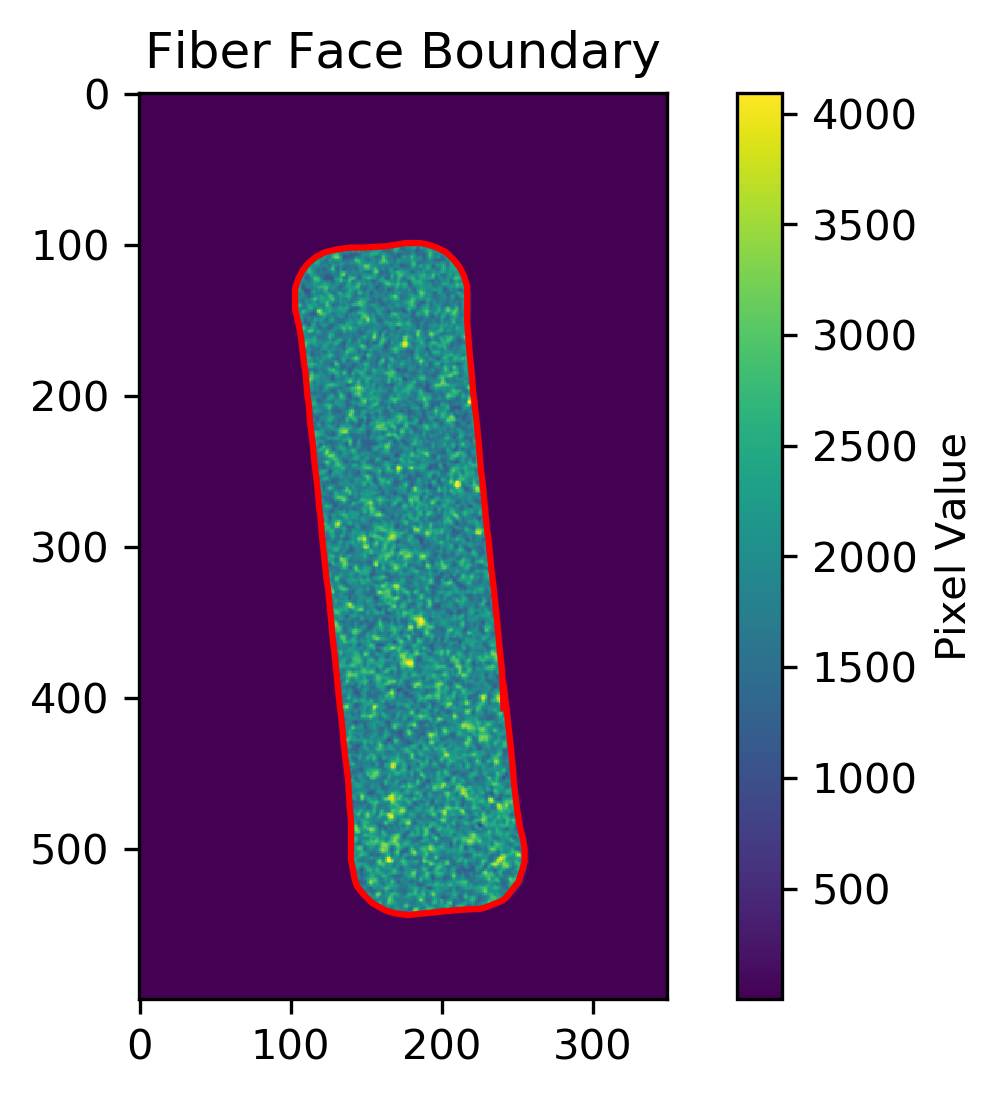}
        \caption{Fiber face boundary}
        \label{fig:MS_analyze_AS_rect_boundary}
    \end{subfigure}
    \caption{Finding the fiber face boundary in a near-field image of a 400 \textmu m by 100 \textmu m rectangular fiber (Figure \ref{fig:MS_analyze_AS_rect_og}, taken when the mode scrambler was off), using the procedure described in Section \ref{sec:fiber_boundary}. Figure \ref{fig:MS_analyze_AS_rect_boundary} shows the boundary in red. Note that the boundary for this fiber is concave. The axes represent pixel number.}
    \label{fig:MS_analyze_AS_rect}
\end{figure}

To find the alpha shape of $S$ given some $\alpha$, firstly the Delaunay triangulation of $S$, called $\textrm{DT}(S)$, was found. This was done using SciPy\cite{2020SciPy-NMeth}. Next, the alpha complex $\mathcal{C}_\alpha(S)$ was found. $\mathcal{C}_\alpha(S)$ is a subcomplex of $\textrm{DT}(S)$ and consists of all simplices in $\textrm{DT}(S)$ such that the radius of the circumcircle of the simplex is less than $\alpha$ \cite{Fischer2000}. Finally, the boundary of $\mathcal{C}_\alpha(S)$ is the boundary of the alpha shape. This boundary is made up of a set of lines represented by the lines' endpoints. However, note that there can be both an inner boundary and an outer boundary. The outer boundary is desired, and it is assumed that the outer boundary is the boundary which has more points. Figures \ref{fig:MS_analyze_AS_Oct}, \ref{fig:MS_analyze_AS_Sq}, and \ref{fig:MS_analyze_AS_rect} show this procedure applied to a near-field images of an octagonal, square, and rectangular fiber respectively. $\alpha=100$ was used for these examples. In particular, notice that the rectangular fiber does not have a convex boundary, which is why this procedure was needed.

\begin{figure}[H]
    \centering
    \includegraphics[width=0.9\textwidth]{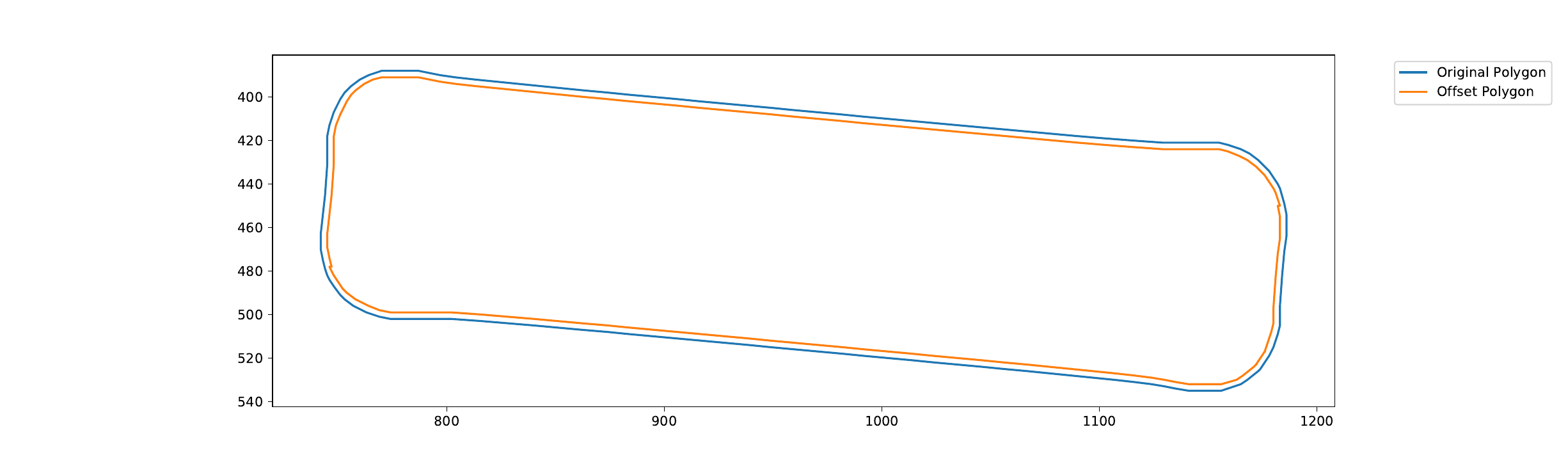}
    \caption{An example of a boundary being offset by 3 pixels, for a rectangular fiber. The original boundary is shown in blue, and the offset boundary is shown in orange. The axes represent pixel number.}
    \label{fig:MS_analyze_rect_offset}
\end{figure}

Finally, the alpha shape was offset inwards by a certain number of pixels in order to account for the actual physical boundary of the fiber between the cladding and core. This was done using the Python pyclipper library, which is a wrapper of the C++ Clipper library\cite{Clipper}. The boundary of this offset shape was taken as the boundary of the fiber face. An example of this is shown in Figure \ref{fig:MS_analyze_rect_offset}, for a rectangular fiber. With the fiber face boundary identified, metrics for scrambling can be computed from the pixels inside the boundary.

\subsection{Metric for Mode Scrambling}\label{sec:metric_SNR}

The metric we used to describe the amount of mode scrambling was the ``signal-to-noise ratio'' (SNR), which we defined as:
\begin{equation}
    \textrm{SNR} \equiv \frac{\mu}{\sigma}
\end{equation}
where $\mu$ and $\sigma$ are:
\begin{align}
    \mu &= \frac{1}{N} \sum_{i=1}^{N} x_i \\
    \sigma &= \sqrt{\frac{1}{N} \sum_{i=1}^{N} (x_i - \mu)^2}
\end{align}
where $N$ is the total number of pixels inside the fiber face boundary and each $x_i$ is the value of one pixel. $\mu$ is the mean pixel value for the pixels inside the fiber face boundary, and $\sigma$ is the standard deviation of the pixels inside the fiber face boundary. This definition was inspired by a metric used in Ref. \citenum{Petersburg2018} for mode scrambling. In Ref. \citenum{Petersburg2018}, a quantity called the ``signal-to-noise ratio'' was also defined as a metric for mode scrambling, but is slightly different from the definition here in that the median is used instead of the mean and the image was firstly median filtered. The uncertainty in $\mu$, $\sigma$, and SNR are $\delta \mu$, $\delta\sigma$, and $\delta \textrm{SNR}$ respectively, given as:
\begin{align}
    \delta \mu &= \frac{1}{N} \sqrt{\sum_{i=1}^{N} (\delta x_i)^2} \\
    \delta \sigma &= \sqrt{\sum_{i=1}^N \left[ \left( \frac{x_i - \mu}{\sigma N} \right)^2 (\delta x_i)^2 \right]}\\
    \delta \textrm{SNR} &= (\textrm{SNR}) \sqrt{\left( \frac{\delta \mu}{\mu}\right)^2 + \left( \frac{\delta \sigma}{\sigma}\right)^2}
\end{align}
\section{Results and Discussion}\label{sec:results}

\begin{figure}[H]
    \centering
    \begin{subfigure}{0.38\textwidth}
        \centering
        \includegraphics[width=\textwidth]{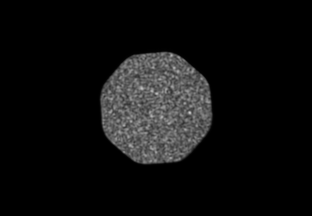}
        \caption{Mode scrambler off; $\textrm{SNR} \approx 4.20$}
        \label{fig:MS_analyze_ex_img_oct_off}
    \end{subfigure}
    \begin{subfigure}{0.38\textwidth}
        \centering
        \includegraphics[width=\textwidth]{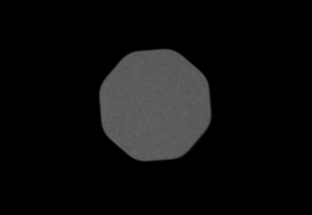}
        \caption{Mode scrambler on; $\textrm{SNR} \approx 32.00$}
        \label{fig:MS_analyze_ex_img_oct_on}
    \end{subfigure}
    \caption{Near-field images of a 100 \textmu m octagonal test fiber, with the mode scrambler was off (Figure \ref{fig:MS_analyze_ex_img_oct_off}) and on (Figure \ref{fig:MS_analyze_ex_img_oct_on}). Both images are a stack of 100 frames, with each individual frame having an exposure time of 60 \textmu s. Both images were acquired when the four-bar linkage crank-rocker (Figure \ref{fig:four-bar_lego}) was used, and bar $AM$ was driven at 5 Hz.}
    \label{fig:MS_analyze_ex_img_oct}
\end{figure}

\begin{figure}[H]
    \centering
    \begin{subfigure}{0.38\textwidth}
        \centering
        \includegraphics[width=\textwidth]{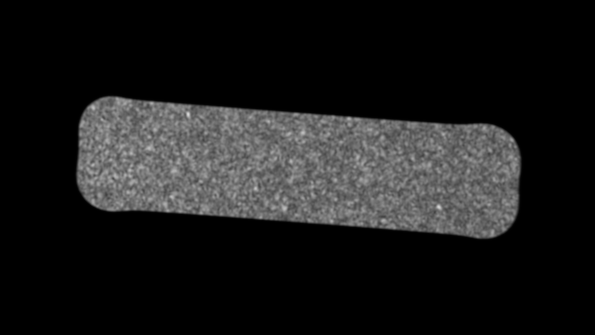}
        \caption{Mode scrambler off; $\textrm{SNR} \approx 5.52$}
        \label{fig:MS_analyze_ex_img_rect_off}
    \end{subfigure}
    \begin{subfigure}{0.38\textwidth}
        \centering
        \includegraphics[width=\textwidth]{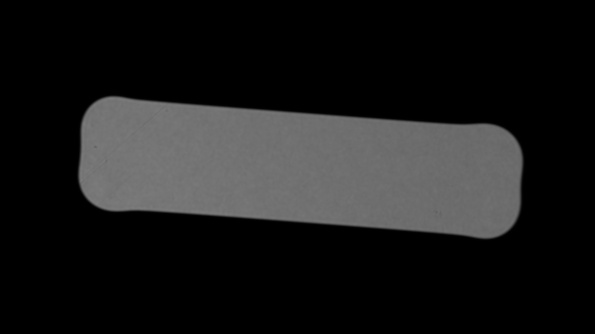}
        \caption{Mode scrambler on; $\textrm{SNR} \approx 35.68$}
        \label{fig:MS_analyze_ex_img_rect_on}
    \end{subfigure}
    \caption{Near-field images of a 400 \textmu m by 100 \textmu m rectangular test fiber, with the mode scrambler was off (Figure \ref{fig:MS_analyze_ex_img_rect_off}) and on (Figure \ref{fig:MS_analyze_ex_img_rect_on}). Both images are a stack of 100 frames, with each individual frame having an exposure time of \mbox{350 \textmu s}. Both images were acquired when the crank-rocker (Figure \ref{fig:four-bar_lego}) was used, and bar $AM$ was driven at 5 Hz.}
    \label{fig:MS_analyze_ex_img_rect}
\end{figure}

In this section, we discuss the results of some scrambling experiments we conducted using our mode scrambler designs described in Section \ref{sec:MS_design}. We tested our designs on several step-index multimode optical fibers with octagonal, square, and rectangular core cross-sections, and varied parameters including shaft distance and agitation frequency. We also investigated the effects of stacking images, different exposure times, and looping the fiber through a shaft with holes (Figure \ref{fig:single_arm}). We took images of test fibers using the near-field arm of our FCS. In all of the following experiments, we used a 635 nm laser diode as the light source (5 \textmu m single mode fiber feed at pre-injection arm). Figures \ref{fig:MS_analyze_ex_img_oct} and \ref{fig:MS_analyze_ex_img_rect} show examples of images of the near-field of an octagonal and rectangular fiber respectively, when the crank-rocker mode scrambler (Figure \ref{fig:four-bar_lego}) was off and on. Modal noise was reduced by around a factor of 8.

\subsection{Exposure Time and Number of Stacked Frames}\label{sec:results_exp_time_stacked_frames}

The first set of experiments involved investigating the effects of exposure time and the number of stacked frames on scrambling. Intuitively, an image taken with a longer exposure time or a stack of images will appear more scrambled because the modal noise pattern will change over time. However, one cannot simply say that increasing the exposure time or number of stacked frames constitutes a mode scrambler. There should be some upper limit for both the exposure time and number of stacked frames, beyond which scrambling will only improve marginally. These limits are due to the mode scrambler's inherent agitation performance. We first investigated these limits before varying other mode scrambler parameters.

Three sets of data were taken in order to see the effect of exposure time and number of stacked frames on scrambling. This was done using the single rotating arm design (Figure \ref{fig:single_arm}). In all of these three experiments, the fiber was positioned at $7''$ away from the motor shaft. When the mode scrambler was on, the period was held constant at 2 s. The test fiber was a 150 \textmu m square fiber.

\subsubsection{Varying Number of Stacked Frames}

\begin{figure}[h]
    \centering
    \includegraphics[width=0.63\textwidth]{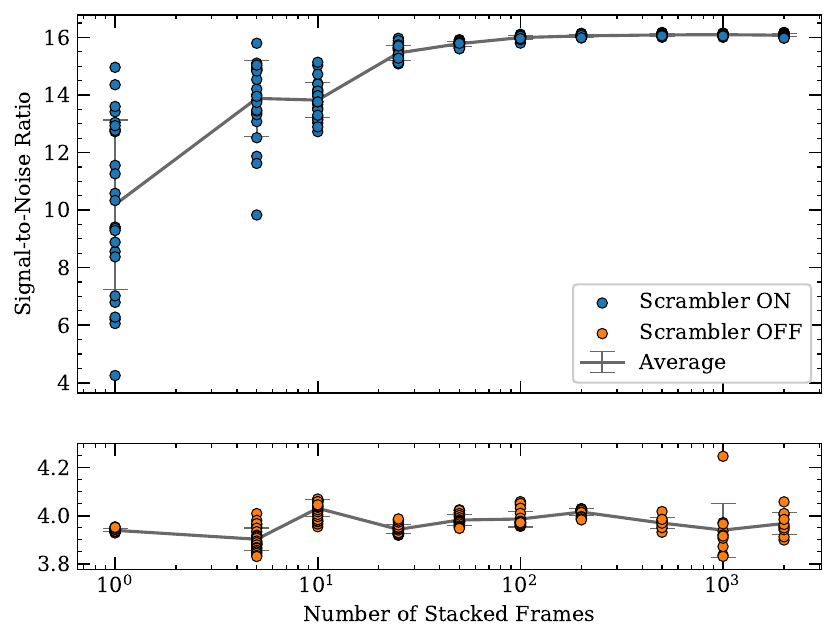}
    \caption{Plot of SNR versus number of stacked frames. Each individual frame had an exposure time of 750 \textmu s. The test fiber used was a 150 \textmu m square fiber.}
    \label{fig:MS_analyze_stack_frame_vary}
\end{figure}

In this experiment, the number of stacked frames was varied between 1 and 2000, while the exposure time of individual frames was held constant at 750 \textmu s. The result is shown in Figure \ref{fig:MS_analyze_stack_frame_vary}. In this figure, each data point represents a image stack consisting of the number of frames indicated on the horizontal axes. The individual frames in the stack all have an exposure time of 750 \textmu s and were acquired at 25 FPS. In general, as the number of stacked frames increases, SNR also increases. However, there is a plateau in the data. The SNR values for $\gtrsim100$ stacked frames are similar, which suggests that there is a certain point where increasing the number of stacked frames would not improve scrambling significantly.


\subsubsection{Varying Exposure Time of Individual Frames}

\begin{figure}[h]
    \centering
    \includegraphics[width=0.6\textwidth]{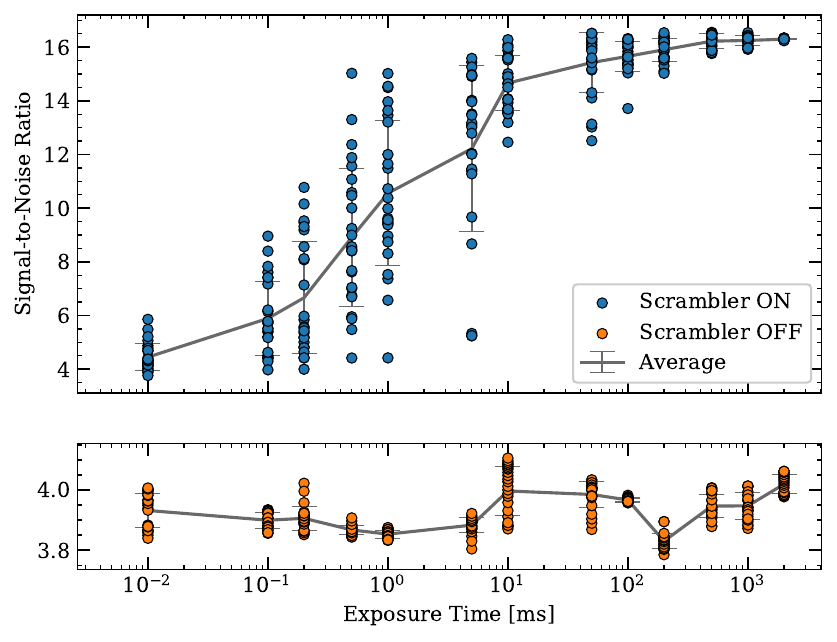}
    \caption{Plot of SNR versus exposure time for individual frames. The test fiber used was a 150 \textmu m square fiber.}
    \label{fig:MS_analyze_exp_vary}
\end{figure}

In this experiment, no frames were stacked, and the exposure time was varied between 10 \textmu s (the minimum possible exposure time for the Matrix Vision mvBlueCOUGAR-XD camera) and \mbox{2 s}. The result is shown in Figure \ref{fig:MS_analyze_exp_vary}. For each exposure time, 25 frames were acquired at a limiting frame rate of 25 FPS. In general, as the exposure time increases, SNR also increases. However, the range of the average SNR values when varying exposure time ($\sim$4 to $\sim$16 in Figure \ref{fig:MS_analyze_exp_vary}) appears to be greater than the range of the average SNR values when varying the number of stacked frames ($\sim$10 to $\sim$16 in Figure \ref{fig:MS_analyze_stack_frame_vary}), which suggests that perhaps exposure time has a more significant affect on SNR. In addition, there again appears to be a plateau in the data. There is not much difference in SNR between an exposure time of 0.5 s and 2 s.


\subsubsection{Constant Total Exposure Time}

\begin{figure}[h]
    \centering
    \includegraphics[width=0.63\textwidth]{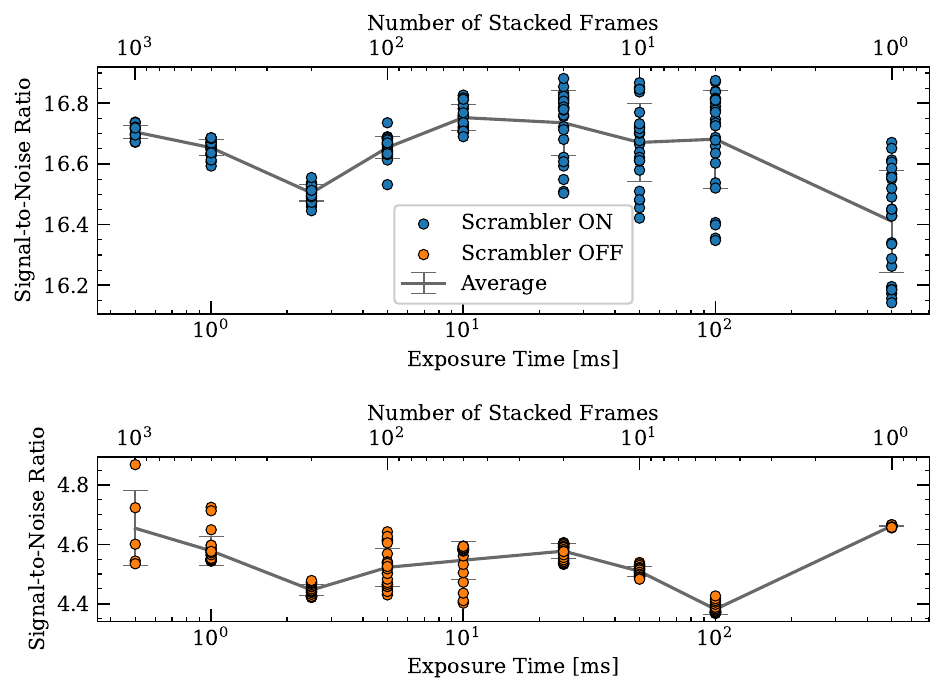}
    \caption{SNR for a constant total exposure time of 0.5 s, while varying individual frame exposure time and the number of stacked frames. The test fiber used was a 150 \textmu m square fiber.}
    \label{fig:MS_analyze_tot_exp_const}
\end{figure}

In this experiment, the total exposure time was held constant at 0.5 s, and the exposure times of individual frames and number of stacked frames were varied. That is, the product of the individual frame exposure time and the number of stacked frames was held constant. The result is shown in Figure \ref{fig:MS_analyze_tot_exp_const}. In this figure, each data point represents a stacked image consisting of $N_\textrm{stack}$ frames, with each individual frame having an exposure time of $t_{\textrm{individual}}$, where $N_\textrm{stack}$ and $t_{\textrm{individual}}$ are the values on the upper and lower horizontal axes respectively. Frames were acquired at a limiting frame rate of 25 FPS. While keeping the total exposure time constant, the SNR did not change as significantly when compared to the previous two experiments. Read noise might be a factor at play here for the shorter exposures. Exposure time may have a more significant effect on scrambling than just stacking frames. 


\subsection{Shaft Distance}\label{sec:results_dist_vary}

\begin{figure}[h]
    \centering
    \includegraphics[width=0.6\textwidth]{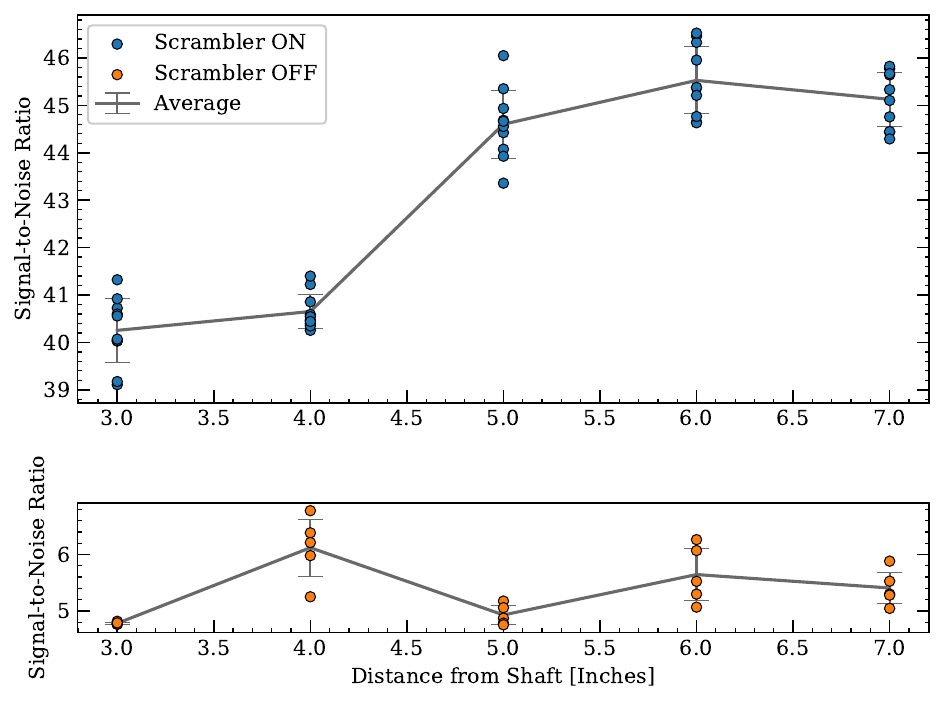}
    \caption{Plot of SNR versus distance away from shaft, for a freely-moving 400 \textmu m by \mbox{100 \textmu m} rectangular fiber}
    \label{fig:MS_analyze_dist_vary}
\end{figure}

In this experiment, the distance between the fiber and the motor shaft was varied between $3''$ and $7''$. The single rotating arm design (Figure \ref{fig:single_arm}) was used, and the period was held constant at \mbox{2 s}. The result is shown in Figure \ref{fig:MS_analyze_dist_vary}. In this figure, each data point represents a stacked image consisting of 100 frames, with each individual frame having an exposure time of 650 \textmu s. A 400 \textmu m by \mbox{100 \textmu m} rectangular fiber was used as the test fiber. From the figure, one can see that for the most part, as the distance between the fiber and the motor shaft increases, the SNR also increases. The positive correlation between these two variables is intuitive because if the fiber is further away from the motor shaft, then it will move more and hence be agitated more. However, note that there appears to be some sort of plateau for distances beyond $6''$. It could be the case that if the distance is large, increasing it further would not make much of a difference on scrambling. More investigation is required.


\subsection{Looping through Several Holes}\label{sec:results_loop}

\begin{figure}[h]
    \centering
    \includegraphics[width=0.48\textwidth]{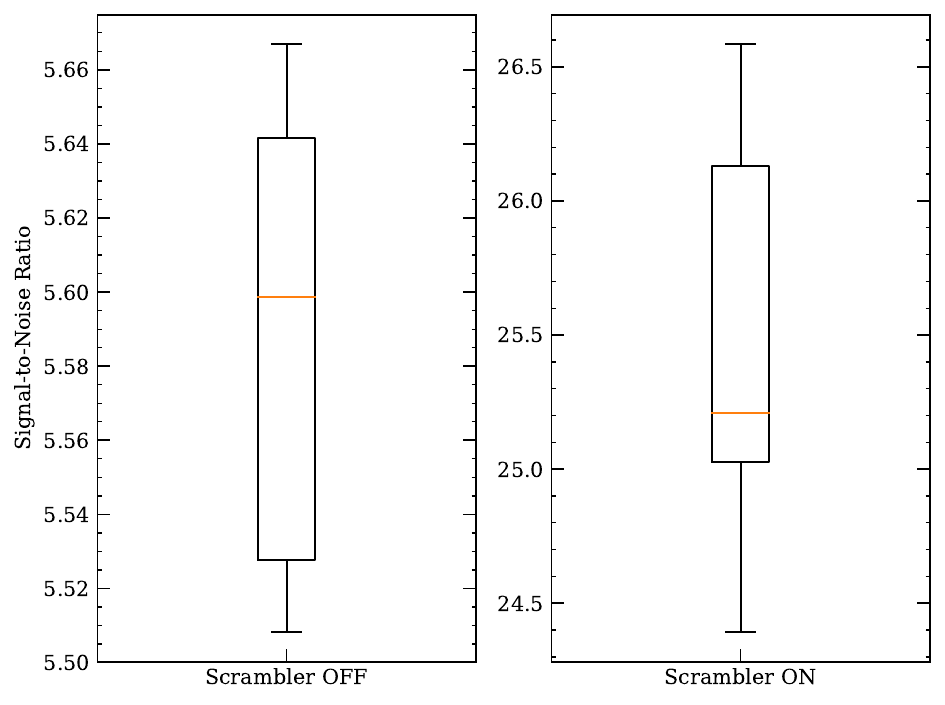}
    \caption{SNR for configuration where a test fiber is looped through several holes in the rotating arm. The test fiber used was a 400 \textmu m by 100 \textmu m rectangular fiber.}
    \label{fig:MS_analyze_loop}
\end{figure}

In this experiment, a rectangular fiber was looped/laced through all five holes in the single rotating arm design (Figure \ref{fig:single_arm}). The period of motor was held constant again at \mbox{2 s}. The result is shown in the box-and-whisker plots in Figure \ref{fig:MS_analyze_loop}. The left box-and-whisker plot contains data from 5 stacked frames, each consisting of 100 individual frames taken when the mode scrambler was off. The right box-and-whisker plot contains data from 20 stacked frames, each consisting of 100 individual frames taken when the mode scrambler was on. All of the individual frames had an exposure time of 500 \textmu s.

From the figure, one can see that there is some scrambling going on when the mode scrambler is turned on. However, the SNR for when the mode scrambler is turned on here is less than that of in Figure \ref{fig:MS_analyze_dist_vary}, when the test fiber was just passed through one hole. One could say that for the rectangular fiber, not looping/lacing through all the holes is better than looping/lacing through all the holes. This may be due to the stiffness of our rectangular fiber's jacket. When the rectangular fiber is looped/laced through all the holes, the parts of the fiber between the holes cannot move too much in comparison to when the fiber is passed through only one hole. There is less agitation when the fiber is looped/laced through all the holes.


\subsection{Frequency}\label{sec:results_frequency}

In this experiment, we varied the motor frequency of the four-bar linkage crank-rocker design (Figure \ref{fig:four-bar_lego}). The motor frequency was varied from 0.1 Hz to 3 Hz. Since there is a 40:24 gear ratio, the actual frequency of bar $AM$ (see Figure \ref{fig:four-bar}) varies from $\sim$0.17 Hz to 5 Hz. The frequency of bar $AM$ is shown in the following plots.

\begin{figure}[h]
    \centering
    \includegraphics[width=0.63\textwidth]{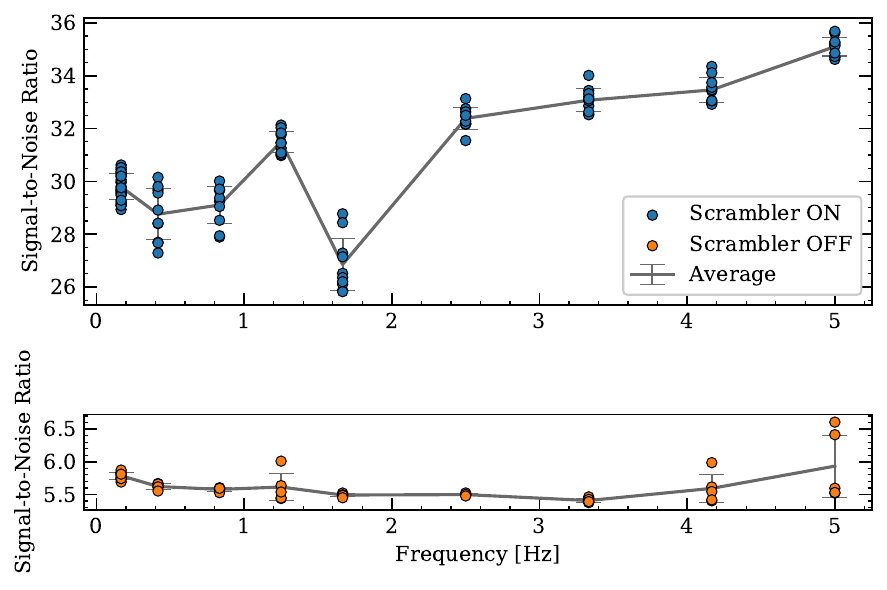}
    \caption{Plot of SNR versus frequency for a 400 \textmu m by 100 \textmu m rectangular fiber. Each data point represents 100 stacked frames, with each individual frame having an exposure time of 350 \textmu s.}
    \label{fig:MS_analyze_freq_rect}
\end{figure}

\begin{figure}[h]
    \centering
    \includegraphics[width=0.63\textwidth]{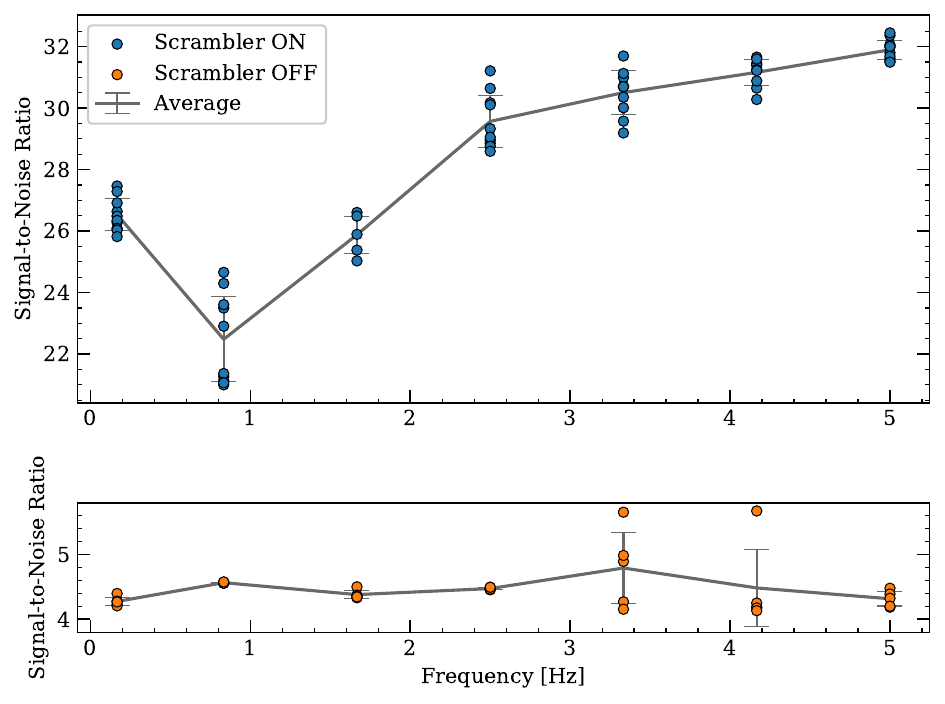}
    \caption{Plot of SNR versus frequency for a 100 \textmu m octagonal fiber. Each data point represents 100 stacked frames, with each individual frame having an exposure time of 60 \textmu s.}
    \label{fig:MS_analyze_freq_oct}
\end{figure}

Figure \ref{fig:MS_analyze_freq_rect} shows a plot of SNR versus frequency for a rectangular test fiber. Generally, as the frequency of the bar increases, the SNR also increases. The fiber is agitated more with higher frequencies, and hence there is better scrambling. Figure \ref{fig:MS_analyze_freq_oct} shows a plot of SNR versus frequency for an octagonal test fiber. Like in Figure \ref{fig:MS_analyze_freq_rect}, as the frequency of the bar increases, the SNR also increases. In both Figures \ref{fig:MS_analyze_freq_rect} and \ref{fig:MS_analyze_freq_oct}, it appears that the SNR will continue to increase with frequencies greater than 5 Hz. However, we limited  the frequency of bar $AM$ to 5 Hz in order to avoid damage to the fibers. More investigation could be conducted into what the maximum frequency should be. Comparing Figures \ref{fig:MS_analyze_freq_oct} and \ref{fig:MS_analyze_freq_rect}, the rectangular fiber appears to have slightly better mode scrambling than the octagonal fiber, even when the mode scrambler was off. This is consistent with results in the literature which show that rectangular fibers have better mode scrambling compared to octagonal fibers of similar size\cite{Petersburg2018}, because they can support more modes\cite{Nikitin2011}.

\section{Conclusions and Next Steps}\label{sec:conclusions}

We developed a fiber characterization station (FCS) at SAO CDP, which allows for light to be injected into a test fiber, for images of the fiber near-field, far-field, and injection location to be acquired, and for fiber throughput and focal ratio degradation to be measured. Our FCS consists of two configurations (Sections \ref{sec:FCS_config1} and \ref{sec:FCS_config2}) and four arms: one to inject light into a test fiber (pre-injection arm, Section \ref{sec:pre-injection_arm}), one to image the fiber injection position (injection imaging arm, Section \ref{sec:injection_imaging_arm}), one to image the fiber near-field (near-field arm, Section \ref{sec:near_field_arm}), and one to image the fiber far-field (far-field arm, Section \ref{sec:far_field_arm}). This FCS was used to test our mode scrambler. While our FCS worked well for our experiments, the $90^\circ$ OAPs were somewhat difficult to align for inexperienced users, although alignment was successfully completed using a systematic procedure. As a next step, we are currently developing a FCS which is easier to align, by relying less on OAPs and using $45^\circ$ OAPs instead, and also improves on the ability to measure scrambling gain. This newer FCS design is presented in Appendix \ref{sec:new_FCS}.

We developed a prototype optical fiber mode scrambler for G-CLEF, which uses a four-bar linkage crank-rocker design (Figure \ref{fig:four-bar_lego}). This design was an extension and improvement of a simpler design using one single rotating arm (Figure \ref{fig:single_arm}). We tested these designs on several step-index multimode optical fibers with octagonal, square, and rectangular core cross-sections. We developed an image analysis procedure utilizing alpha shapes to identify the boundary of the face of a fiber (Section \ref{sec:fiber_boundary}) and then computed a signal-to-noise ratio (SNR) metric for mode scrambling (Section \ref{sec:metric_SNR}). From the mode scrambler experiments, increasing the exposure time and number of stacked frames increases SNR, but only up to a certain point (Section \ref{sec:results_exp_time_stacked_frames}). Increasing the distance between the test fiber and the motor shaft increases SNR (Section \ref{sec:results_dist_vary}), but for larger distances, perhaps the effect becomes negligible. Looping/lacing a rectangular fiber through all holes in the single rotating arm design makes the scrambling performance worse than if the fiber just passes through one hole (Section \ref{sec:results_loop}). Increasing the motor frequency increases SNR (Section \ref{sec:results_frequency}). The rectangular fiber appears to have slightly better mode scrambling than the octagonal fiber. Some next steps would be to conduct more experiments using the four-bar linkage crank-rocker design and to investigate a wider variety of optical fiber cross-sections and sizes. Another idea is to investigate the effect of using multiple mechanisms to agitate the fiber, such as a series of crank-rockers operating at different frequencies. Ultimately, this work has demonstrated a successful prototype mode scrambler, investigated the effects of different agitation parameters, and established a robust experimental framework for future investigations into mode scrambling for precision RV spectrographs. 
\appendix
\section{Next-Generation Fiber Characterization Station (FCS-II)}\label{sec:new_FCS}

\subsection{Design Drivers}

We are developing a new Fiber Characterization Station, referred to as ``FCS-II'', which will be an upgraded version of our FCS described in Section \ref{sec:FCS}. Our FCS-II has three key design drivers:
\begin{enumerate}
    \item Broad-band achromatism (350–950 nm): Coverage spans the full science range of both G@M (390–950~nm) and G-CLEF at the GMT (350–950~nm).
    \item High image quality: Diffraction-limited performance is required when inspecting fibers with core diameters down to 25 \textmu m; this is also crucial for accurate mode scrambling measurements.
    \item Streamlined alignment: The former two-relay with 90$^\circ$ OAPs was difficult to align for inexperienced users; our new FCS-II will be more user-friendly and repeatable.
\end{enumerate}


\subsection{Functional Overview}

\begin{figure}[H]
    \centering
    \includegraphics[width=\textwidth]{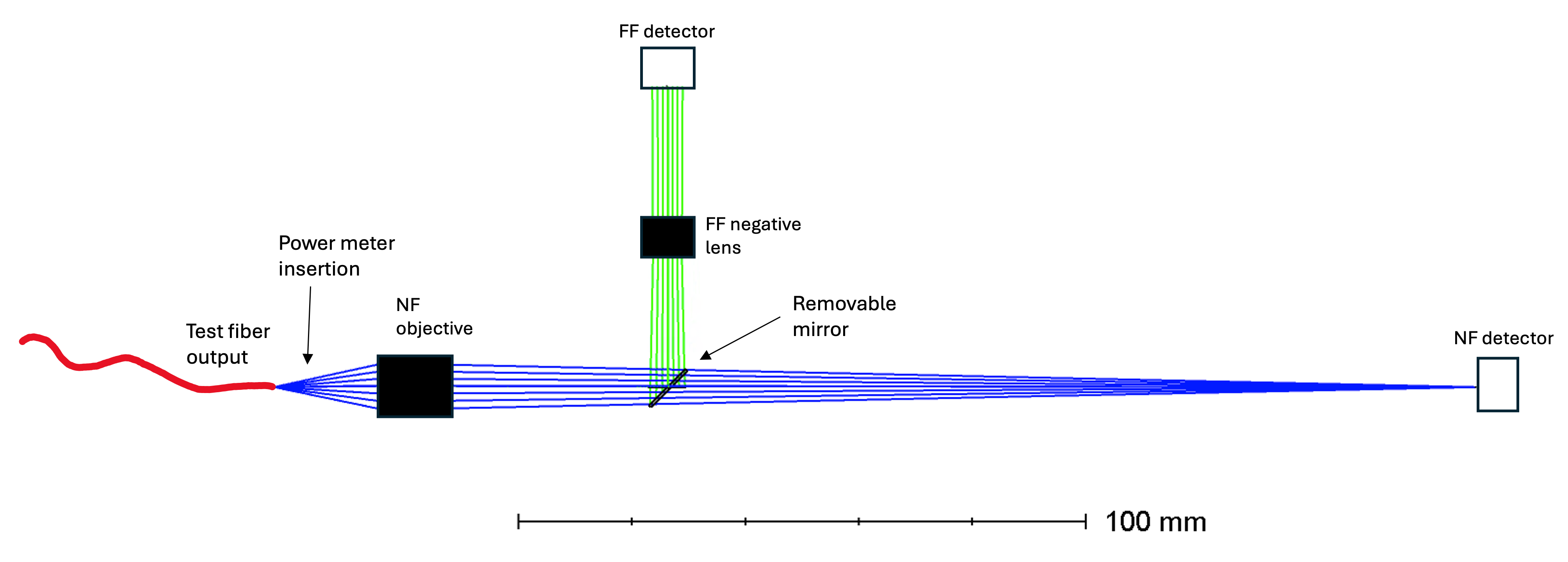}
    \caption{FCS-II inspection arm}
    \label{fig:FCS_II_inspection}
\end{figure}

FCS-II will still provide the four fundamental measurements described in Section \ref{sec:FCS_overview}—mode scrambling, near-field scrambling gain, focal ratio degradation (FRD), and absolute throughput—but will implement new optics to meet the design drivers above. The main changes are listed in Table \ref{tab:FCSII}. Figures \ref{fig:FCS_II_injection} and \ref{fig:FCS_II_inspection} illustrate our FCS-II design.

\begin{table}[H]
\centering
\begin{tabular}{ | m{3cm} | m{4cm}| m{8cm} | } 
  \hline
  Arm/Channel & Key change in FCS-II & Rationale \\ 
  \hline
  \hline
  Injection & Two off-the-shelf 45$^\circ$ OAPs in a 2:1 demagnifying relay & Shallow off-axis angle suppresses coma during deliberate decentering; all-reflective path removes chromatic focus shift; 2:1 ratio produces a fast, compact spot. \\ 
  \hline
  Near-field (pre/post) & Single microscope objective & Only one optical element. Detector is positioned on Z-stage to fine tune focus at band edges.\\ 
  \hline
  Far-field/FRD & Same microscope objective followed by a weak negative lens to collimate & Removes an entire relay, simplifies alignment, and images the far-field pattern onto a 2.5 \textmu m-pixel detector without the need to manually closing an iris. \\
  \hline
  Throughput & Power-meter measurement before and after the test fiber & No crossover optics required; combining absolute throughput with the FRD curve yields a complete loss budget.\\
  \hline
\end{tabular}
\caption{Changes in the FCS-II.}\label{tab:FCSII}
\end{table}


\subsection{Two-stage OAP alignment strategy}

\begin{figure}[H]
    \centering
    \includegraphics[width=\textwidth]{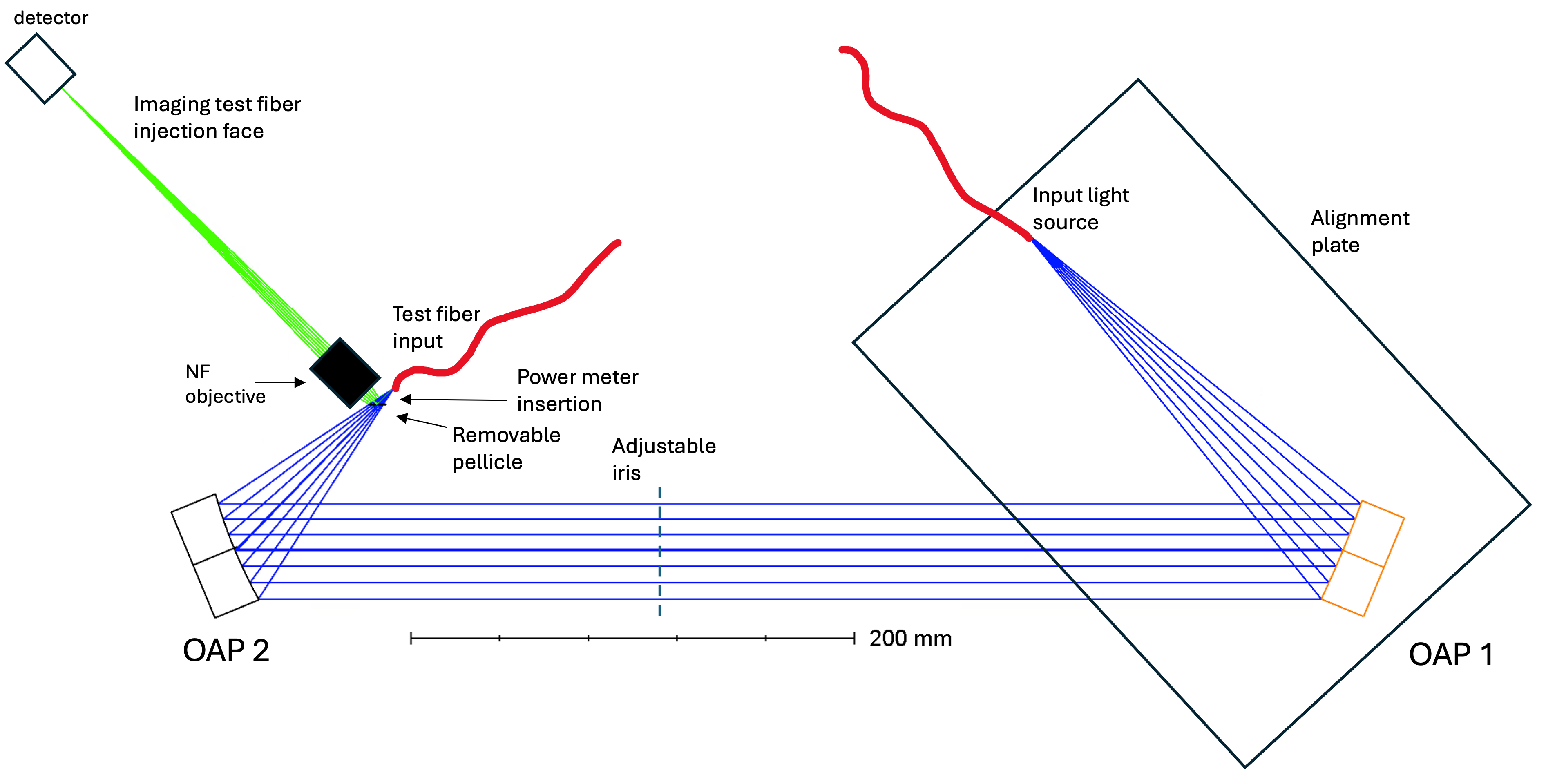}
    \caption{FCS-II injection arm}
    \label{fig:FCS_II_injection}
\end{figure}

Mounting OAPs usually couples tilt and translation, leading to iterative ``focus-then-steer'' loops. We decouple these motions as follows:
\begin{enumerate}
    \item Local fiber registration: Each fiber is centered in the OAP focal spot with an XYZ stage; the fiber holder and OAP are co-mounted on a movable aluminum plate (see Figure \ref{fig:FCS_II_injection}).
    \item Plate matching: The plate is then mated on a common base and adjusted only in tip/tilt about the vertex reference, steering the collimated beam without shifting the focal point, while ensuring the sampling of the chief ray position.
\end{enumerate}

\acknowledgments 
 
M.C.H. Leung thanks the Smithsonian Institution Office of International Relations, Susan Demski-Hamelin, Christine Crowley, and Stuart McMuldroch for organizing the logistics of his internship at the Smithsonian Astrophysical Observatory, which led to this work. \mbox{M.C.H. Leung} gratefully acknowledges the support of the Natural Sciences and Engineering Research Council of Canada (NSERC) through an NSERC Postgraduate Scholarship – Doctoral (PGS D). M.C.H. Leung thanks SPIE for providing a conference fee waiver and travel grant to attend SPIE Optics + Photonics 2025. 

\bibliography{report} 
\bibliographystyle{spiebib} 

\end{document}